\documentclass[12pt]{article}
\usepackage{amssymb}
\usepackage{epsf}
\usepackage{epsfig}
\usepackage{cite}
\newcommand{\lsim}{
\mathrel{\hbox{\rlap{\hbox{\lower4pt\hbox{$\sim$}}}\hbox{$<$}}}}

\newcommand{\gsim}{
\mathrel{\hbox{\rlap{\hbox{\lower4pt\hbox{$\sim$}}}\hbox{$>$}}}}

\begin{document}
\begin{titlepage}
\vspace*{-0.5truecm}

\begin{flushright}
CERN-PH-TH/2007-080\\
\end{flushright}

\vspace*{1.3truecm}

\begin{center}
\boldmath
{\Large{\bf 
$B_{s,d}\to \pi\pi, \pi K, KK$: Status and Prospects
}}
\unboldmath
\end{center}

\vspace{0.9truecm}

\begin{center}
{\bf Robert Fleischer}

\vspace{0.5truecm}

{\sl Theory Division, Department of Physics, CERN, CH-1211 Geneva 23,
Switzerland}

\end{center}

\vspace{0.6cm}
\begin{abstract}
\vspace{0.2cm}\noindent
Several years ago, it was pointed out that the $U$-spin-related decays
$B_d\to\pi^+\pi^-$, $B_s\to K^+K^-$ and $B_d\to\pi^\mp K^\pm$, 
$B_s\to \pi^\pm K^\mp$ offer interesting strategies for the extraction of the angle 
$\gamma$ of the unitarity triangle. Using the first results from the Tevatron on the 
$B_s$ decays and the $B$-factory data on $B_{u,d}$ modes, we compare the 
determinations of $\gamma$ from both strategies, study the sensitivity on 
$U$-spin-breaking effects, discuss the resolution of discrete ambiguities, predict 
observables that were not yet measured but will be accessible at LHCb, explore the 
extraction of the width difference $\Delta\Gamma_s$ from untagged $B_s\to K^+K^-$
rates, and address the impact of new physics. The data for the  $B_d\to\pi^+\pi^-$, 
$B_s\to K^+K^-$ system favour the BaBar measurement of direct CP violation in 
$B_d\to\pi^+\pi^-$, which will be used in the numerical analysis, and result in a 
fortunate situation, yielding $\gamma=(66.6^{+4.3+4.0}_{-5.0-3.0})^\circ$, where the 
latter errors correspond to a generous estimate of $U$-spin-breaking effects. On the 
other hand, the $B_d\to\pi^\mp K^\pm$, $B_s\to \pi^\pm K^\mp$ analysis leaves us 
with $26^\circ\leq\gamma\leq70^\circ$, and points to a value of the 
$B_s\to \pi^\pm K^\mp$ branching ratio that is larger than the current Tevatron result. 
An important further step will be the measurement of mixing-induced CP violation in 
$B_s\to K^+K^-$, which will also allow us to extract the $B^0_s$--$\bar B^0_s$ mixing 
phase unambiguously with the help of $B_s\to J/\psi \phi$ at the LHC. Finally, the 
measurement of direct CP violation in $B_s\to K^+K^-$ will make the full exploitation 
of the physics potential of the $B_{s,d}\to \pi\pi, \pi K, KK$ modes possible.
\end{abstract}

\vspace*{0.5truecm}
\vfill
\noindent
May 2007

\end{titlepage}

\thispagestyle{empty}
\vbox{}
\newpage

\setcounter{page}{1}

\section{Introduction}\label{sec:intro}
\begin{figure}
\centerline{
 \includegraphics[width=4.8truecm]{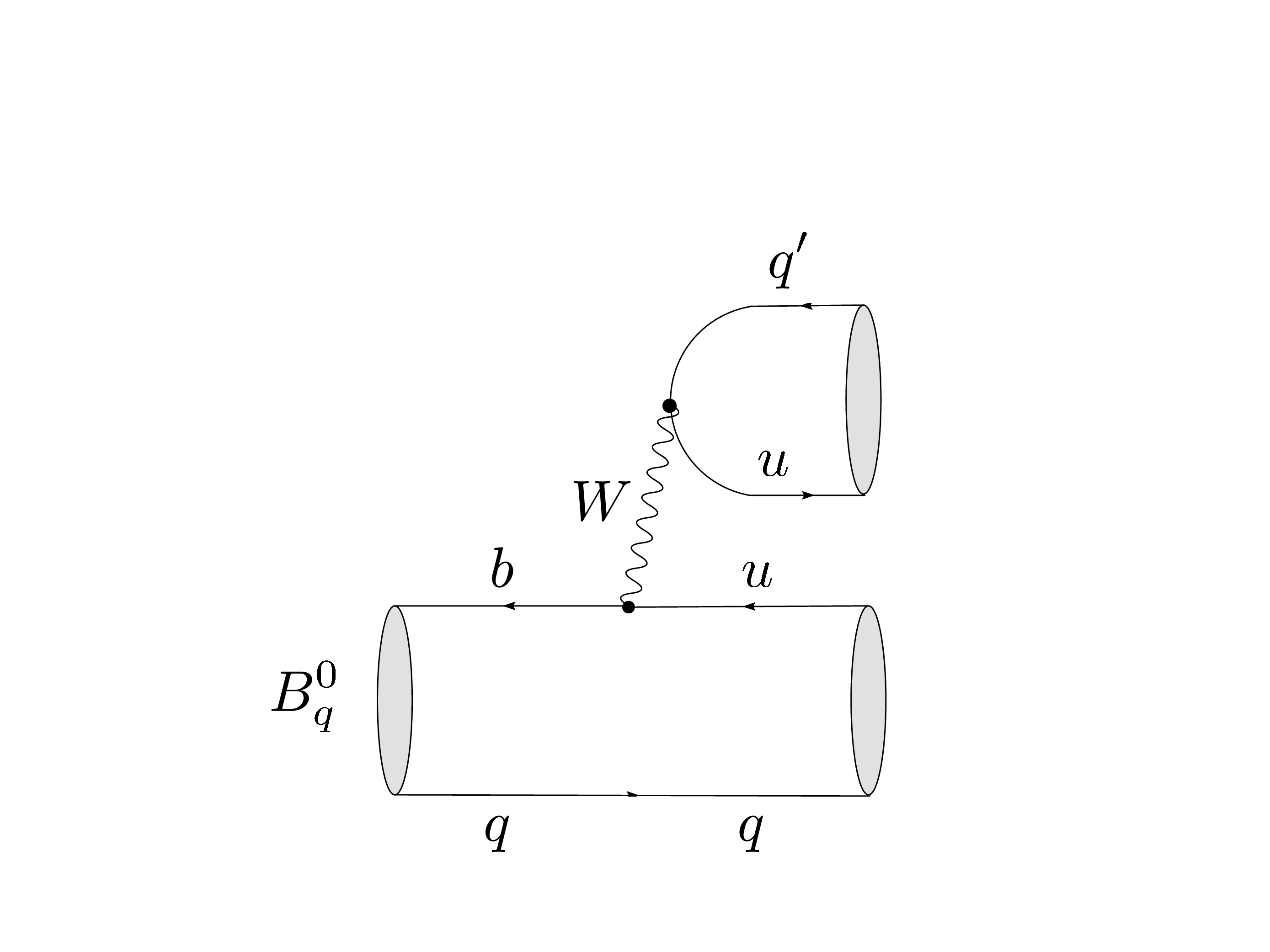}
 \hspace*{0.5truecm}
 \includegraphics[width=5.4truecm]{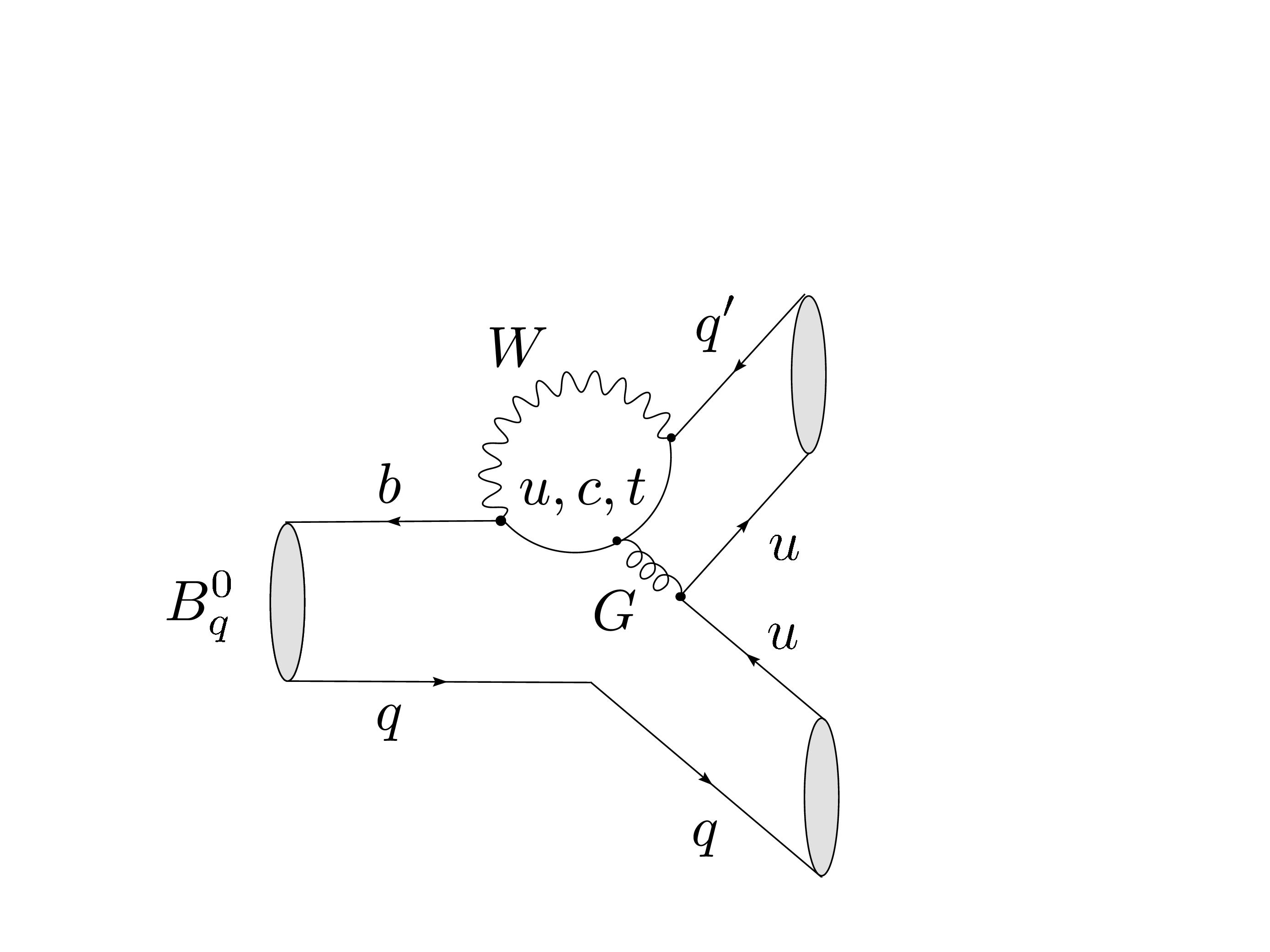}  
 }
 \vspace*{-0.1truecm}
\caption{Tree and penguin topologies contributing to the 
$U$-spin-related $B^0_d\to\pi^+\pi^-$, $B^0_s\to K^+K^-$ and
$B^0_d\to\pi^- K^+$, $B^0_s\to \pi^+K^-$ decays 
($q,q'\in\{d,s\}$).}\label{fig:U-spin-diag}
\end{figure}

\setcounter{equation}{0}
Decays of $B$ mesons into two light pseudoscalar mesons offer interesting
probes for the exploration of CP violation. The key problem in these studies
is usually given by the hadronic matrix elements of local four-quark operators, 
which suffer from large theoretical uncertainties. In 1999 \cite{RF-BsKK}, it was 
pointed that the system of the $B^0_d\to \pi^+\pi^-$ and $B^0_s\to K^+K^-$ decays
is particularly interesting in this respect. These transitions, which receive
contributions from tree and penguin topologies, allow us to determine the 
angle $\gamma$ of the unitarity triangle (UT) of the Cabibbo--Kobayashi--Maskawa
(CKM) matrix \cite{ckm} with the help of the $U$-spin symmetry, which is a 
subgroup of the $SU(3)_{\rm F}$ flavour symmetry of strong interactions, 
connecting the strange and down quarks in the same way through $SU(2)$ 
transformations as the isopsin symmetry connects the up and down quarks.
As can be seen in Fig.~\ref{fig:U-spin-diag}, the $B^0_d\to \pi^+\pi^-$ and 
$B^0_s\to K^+K^-$ modes are related to each other through an interchange 
of all down and strange quarks. Consequently, the $U$-spin flavour symmetry
allows us to derive relations between their hadronic parameters so that the
experimental observables offer sufficient information to extract them and
the UT angle $\gamma$ from the data. The advantage of this $U$-spin strategy 
with respect to the conventional $SU(3)$ flavour-symmetry strategies 
\cite{GHLR} is twofold: 
\begin{itemize}
\item no additional dynamical assumptions such as the neglect of annihilation 
topologies have to be made, which could be spoiled by large rescattering effects;
\item electroweak (EW) penguin contributions, which are not invariant
under the isospin symmetry because of the different up- and down-quark
charges, can be included. 
\end{itemize}
The theoretical accuracy is therefore only limited by non-factorizable 
$U$-spin-breaking effects, as the factorizable corrections can be taken into 
account through appropriate ratios of form factors and decay constants. 
Moreover, we have key relations between certain hadronic parameters, where 
these quantities cancel. Interestingly, also experimental insights into 
$U$-spin-breaking effects can be obtained, which do not indicate any 
anomalous enhancement.

The relevant observables are the CP-averaged branching ratios as well as the 
direct and mixing-induced CP asymmetries ${\cal A}_{\rm CP}^{\rm dir}(B_q\to f)$ 
and ${\cal A}_{\rm CP}^{\rm mix}(B_q\to f)$, respectively, entering the following 
time-dependent rate asymmetries for decays into CP eigenstates~\cite{kitz}:
\begin{eqnarray}
{\cal A}_{\rm CP}(t)&\equiv&\frac{\Gamma(B^0_q(t)\to f)-
\Gamma(\bar B^0_q(t)\to f)}{\Gamma(B^0_q(t)\to f)+
\Gamma(\bar B^0_q(t)\to f)}\nonumber\\
&=&\left[\frac{{\cal A}_{\rm CP}^{\rm dir}(B_q\to f)\,\cos(\Delta M_q t)+
{\cal A}_{\rm CP}^{\rm mix}(B_q\to f)\,\sin(\Delta 
M_q t)}{\cosh(\Delta\Gamma_qt/2)-{\cal A}_{\rm 
\Delta\Gamma}(B_q\to f)\,\sinh(\Delta\Gamma_qt/2)}\right],\label{ACP-t}
\end{eqnarray}
where $\Delta M_q$ and $\Delta\Gamma_q$ are the mass and width
differences of the $B_q$ mass eigenstates, respectively. Throughout this paper, 
we shall apply a sign convention for CP asymmetries that is similar to 
(\ref{ACP-t}), also for the direct CP asymmetries of $B$ decays into 
flavour-specific final states. 

As can be seen in Fig.~\ref{fig:U-spin-diag}, there is yet another pair of
$U$-spin-related $B_{d,s}$ decays that is mediated by the same
quark transitions: $B^0_d\to\pi^- K^+$ and $B^0_s\to \pi^+K^-$. In contrast 
to the $B^0_d\to\pi^+\pi^-$, $B^0_s\to K^+K^-$ system, the final states
are flavour-specific. Consequently, we have to rely on the direct CP-violating
rate asymmetry as no mixing-induced CP violation arises. If additional
information provided by the $B^+\to\pi^+K^0$ channel is used, together with
plausible dynamical assumptions about final-state interaction effects and 
colour-suppressed EW penguin topologies, the $U$-spin-related $B^0_d\to\pi^- K^+$, 
$B^0_s\to \pi^+K^-$ decays also allow the extraction of the CKM angle 
$\gamma$ \cite{GR-U-spin}. 

Thanks to the $e^+e^-$ $B$ factories with the BaBar (SLAC) and Belle (KEK)
experiments, the $B^\pm$ and $B_d$ decays are now experimentally 
well established, with the following CP-averaged branching ratios, as 
compiled by the Heavy Flavour Averaging Group (HFAG) \cite{HFAG}:
\begin{eqnarray}
\mbox{BR}(B_d\to\pi^+\pi^-)&=&(5.16\pm0.22)\times 10^{-6},\label{BR-Bdpipi}\\
\mbox{BR}(B_d\to\pi^\mp K^\pm)&=&(19.4\pm0.6)\times 10^{-6},\\
\mbox{BR}(B^\pm\to\pi^\pm K)&=&(23.1\pm1.0)\times 10^{-6}.\label{BR-BppipK}
\end{eqnarray}
The $B_d\to\pi^\mp K^\pm$ channel led to the observation of direct CP violation in 
the $B$-meson system \cite{CP-B-dir}, where the current HFAG average reads as
\begin{equation}\label{CPdir-BdpimKp}
{\cal A}_{\rm CP}^{\rm dir}(B_d\to\pi^\mp K^\pm)=0.095\pm0.013.
\end{equation}
Concerning the measurements of CP violation in $B^0_d\to\pi^+\pi^-$, the 
BaBar and Belle collaborations agree now perfectly on the mixing-induced CP 
asymmetry:
\begin{equation}\label{ACP-mix-pipi-ex}
{\cal A}_{\rm CP}^{\rm mix}(B_d\to\pi^+\pi^-)=
\left\{\begin{array}{ll}
0.60\pm0.11\pm0.03 & \mbox{(BaBar \cite{BaBar-Bpi+pi-})}\\
0.61\pm0.10\pm0.04 & \mbox{(Belle \cite{Belle-Bpi+pi-}),}
\end{array}\right.
\end{equation}
yielding the average of ${\cal A}_{\rm CP}^{\rm mix}(B_d\to\pi^+\pi^-)=
0.61\pm0.08$ \cite{HFAG}.
On the other hand, the picture of direct CP violation is still not experimentally settled,
and the corresponding $B$-factory measurements differ at the $2.6\,\sigma$ level:
\begin{equation}\label{ACP-dir-pipi-ex}
{\cal A}_{\rm CP}^{\rm dir}(B_d\to\pi^+\pi^-)=\left\{
\begin{array}{cc}
-0.21\pm0.09\pm0.02 & \mbox{(BaBar \cite{BaBar-Bpi+pi-})}\\
-0.55\pm0.08\pm0.05 & \mbox{(Belle \cite{Belle-Bpi+pi-}).}
\end{array}\right.
\end{equation}
In a recent paper \cite{FRS}, it was pointed out that the branching ratio and 
direct CP asymmetry of the $B^0_d\to\pi^-K^+$ mode favour actually the
BaBar result. Following a different avenue, we will arrive at the same
conclusion. 

Since the $e^+e^-$ $B$ factories are operated at the $\Upsilon(4S)$ 
resonance, $B_s$ decays could not be studied at these colliders.\footnote{Recently, 
data were taken by Belle at $\Upsilon(5S)$, allowing also access to $B_s$ 
decays \cite{Belle-U5S}.} The exploration of the $B_s$ system is the territory of 
hadron colliders, i.e.\ of the Tevatron (FNAL), which is currently taking data, and 
of the LHC (CERN), which will start operation soon. In fact, signals for 
the $B^0_s\to K^+K^-$ and $B^0_s\to \pi^+K^-$ decays were recently observed 
at the Tevatron by the CDF collaboration at the $4\,\sigma$ and $5\,\sigma$ levels, respectively, which correspond to the following CP-averaged branching ratios:
\cite{CDF-BsK+K-,CDF-punzi}:
\begin{eqnarray}
\mbox{BR}(B_s\to\pi^\pm K^\mp)&=&(5.00\pm0.75\pm1.0)\times 
10^{-6},\label{BR-BspiK}\\
\mbox{BR}(B_s\to K^+K^-)&=&(24.4\pm1.4\pm4.6)\times 10^{-6}.\label{BR-BsKK}
\end{eqnarray}
Moreover, also a CP violation measurement is available:
\begin{equation}\label{ACP-Bs-exp}
{\cal A}_{\rm CP}^{\rm dir}(B_s\to\pi^\pm K^\mp)=-0.39\pm0.15\pm0.08,
\end{equation}
whereas results for the CP-violating observables of $B_s\to K^+K^-$ were not
yet reported.

In view of this progress, it is interesting to confront the $B_d\to\pi^+\pi^-$, 
$B_s\to K^+K^-$ and $B_d\to\pi^\mp K^\pm$, $B_s\to \pi^\pm K^\mp$ strategies
with the measurements performed at the $B$ factories and the Tevatron. 
This is also an important analysis in view of the quickly approaching start of
the LHC with its dedicated $B$-decay experiment LHCb, where the physics potential
of the $B_s$-meson system can be fully exploited \cite{LHCb}. We will therefore
give a detailed presentation, collecting also the relevant formulae, which should
be helpful for the analysis of the future improved experimental data. The 
outline of this paper is as follows: in Section~\ref{sec:BsKK}, we have a closer look 
at the $B_d\to\pi^+\pi^-$, $B_s\to K^+K^-$ strategy, and move on to the 
$B_d\to\pi^\mp K^\pm$, $B_s\to \pi^\pm K^\mp$ system in 
Section~\ref{sec:BspiK}. Finally, we summarize our conclusions in 
Section~\ref{sec:concl}. For analyses using QCD factorization, soft collinear effective
theory or perturbative QCD, the reader is referred to 
Refs.~\cite{QCDF-ana,SCET-ana,PQCD-ana}.

\boldmath
\section{The $B_d\to\pi^+\pi^-$, $B_s\to K^+K^-$ Strategy}\label{sec:BsKK}
\unboldmath
\setcounter{equation}{0}
\boldmath
\subsection{CP Violation in $B_d\to\pi^+\pi^-$}
\begin{figure}
\centerline{
 \includegraphics[width=7.5truecm]{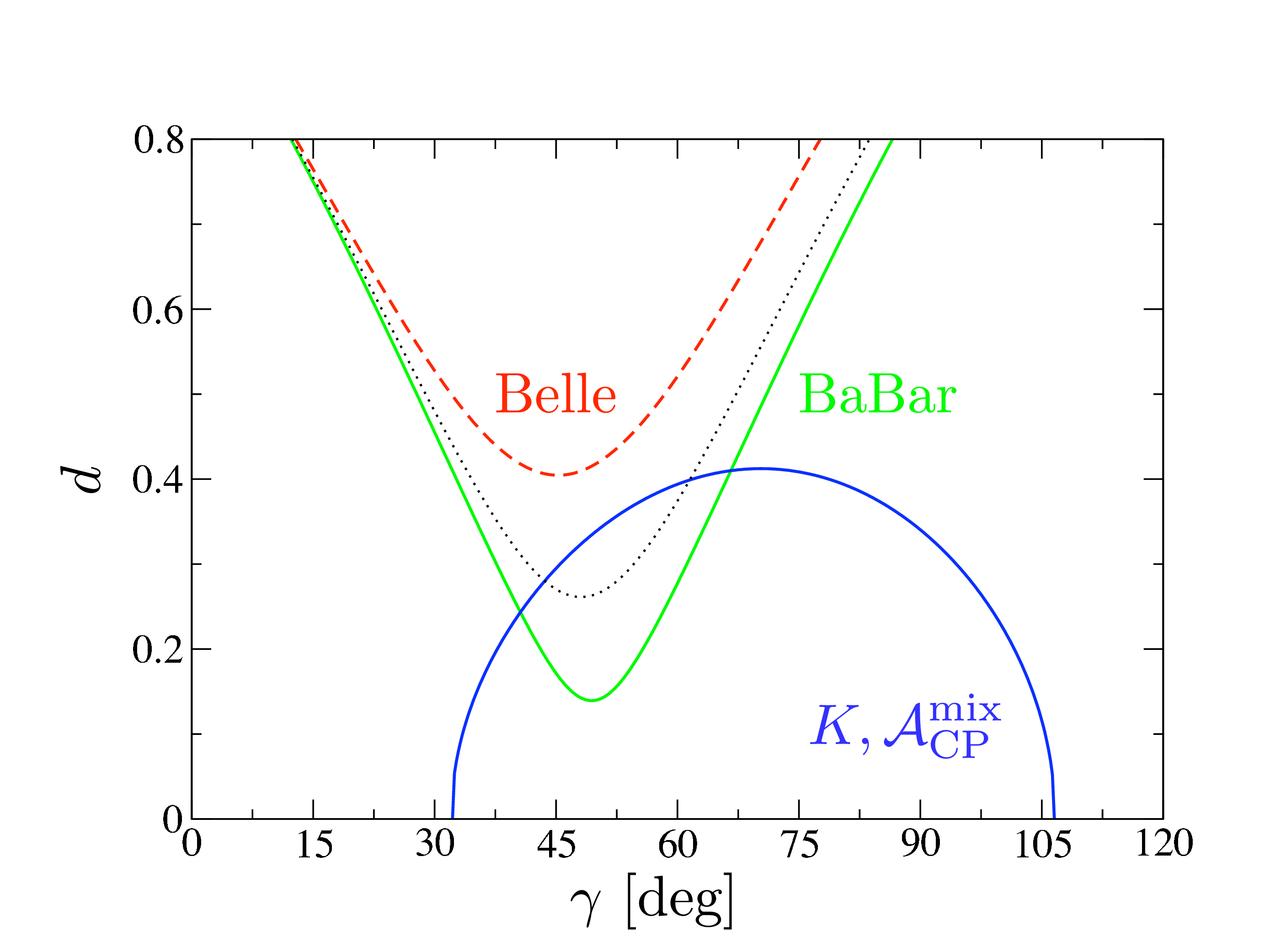}}
 \vspace*{-0.3truecm}
\caption{The contours in the $\gamma$--$d$ plane that follow from the 
central values of the BaBar and Belle measurements of the CP asymmetries 
of the $B_d\to\pi^+\pi^-$ channel and the ratio of the CP-averaged 
$B_d\to\pi^+\pi^-$, $B_s\to K^+K^-$ branching ratios. The dotted line
corresponds to the HFAG average for the direct CP violation in 
$B_d\to\pi^+\pi^-$.}\label{fig:BdpipiBsKK-cont1}
\end{figure}

\unboldmath
In the Standard Model (SM), using the unitarity of the CKM matrix, the
transition amplitude of the $B^0_d\to\pi^+\pi^-$ decay can be written
as follows \cite{RF-BsKK}:
\begin{equation}\label{Bdpipi-ampl}
A(B_d^0\to\pi^+\pi^-)=e^{i\gamma}\left(1-\frac{\lambda^2}{2}\right){\cal C}
\left[1-d\,e^{i\theta}e^{-i\gamma}\right],
\end{equation}
where $\gamma$ is the corresponding angle of the UT, $\lambda$ the 
parameter of the Wolfenstein expansion of the CKM matrix \cite{wolf}, 
${\cal C}$ denotes a CP-conserving strong amplitude that is
governed by the tree contributions, while the CP-consering hadronic 
parameter $d e^{i\theta}$ measures -- sloppily speaking -- the ratio of 
penguin to tree amplitudes. The CP asymmetries introduced in (\ref{ACP-t}) 
take then the following form:
\begin{eqnarray}
{\cal A}_{\rm CP}^{\rm dir}(B_d\to \pi^+\pi^-)&=&
-\left[\frac{2\,d\sin\theta\sin\gamma}{1-
2\,d\cos\theta\cos\gamma+d^2}\right]\label{ACP-dir-d}\\
{\cal A}_{\rm CP}^{\rm mix}(B_d\to \pi^+\pi^-)&=&+\left[\,
\frac{\sin(\phi_d+2\gamma)-2\,d\,\cos\theta\,\sin(\phi_d+\gamma)+
d^2\sin\phi_d}{1-2\,d\cos\theta\cos\gamma+d^2}\,\right],\label{ACP-mix-d}
\end{eqnarray}
where $\phi_d$ is the CP-violating $B^0_d$--$\bar B^0_d$ mixing phase, which
is given by $2\beta$ in the SM, with $\beta$ denoting another UT angle. This
phase has been measured at the $B$ factories with the help of the ``golden"
decay $B^0_d\to J/\psi K_{\rm S}$ and similar modes, including $B_d\to J/\psi K^*$
and $B_d \to D^*D^*K_{\rm S}$ channels to resolve a twofold ambiguity, as
follows \cite{HFAG}:
\begin{equation}
\phi_d=(42.6\pm2)^\circ.
\end{equation} 
The general expressions in (\ref{ACP-dir-d}) and (\ref{ACP-mix-d}) allow us to 
eliminate the strong phase $\theta$, and to calculate $d$ as a function of
$\gamma$ by using the formulae given in Ref.~\cite{RF-BsKK}. In 
Fig.~\ref{fig:BdpipiBsKK-cont1}, we show the corresponding contours for the 
central values of the BaBar and Belle results in (\ref{ACP-mix-pipi-ex}) and 
(\ref{ACP-dir-pipi-ex}). In order to guide the eye, we have also included the
contour (dotted line) representing the central value of the HFAG average 
${\cal A}_{\rm CP}^{\rm dir}(B_d\to\pi^+\pi^-)=-0.38\pm0.07$ of the BaBar
and Belle results for the direct CP violation in $B_d\to\pi^+\pi^-$ \cite{HFAG}. 
It should be emphasized that these contours are valid {\it exaclty} in the SM.

\boldmath
\subsection{CP-Averaged $B_s\to K^+K^-$, $B_d\to\pi^+\pi^-$ Branching Ratios}
\unboldmath
Let now $B^0_s\to K^+K^-$ enter the stage. In analogy to (\ref{Bdpipi-ampl}),
the corresponding decay amplitude can be written as
\begin{equation}\label{BsKK-ampl}
A(B_s^0\to K^+K^-)=e^{i\gamma}\lambda\,{\cal C}'\left[1+\frac{1}{\epsilon}
d'e^{i\theta'}e^{-i\gamma}\right],
\end{equation}
where 
\begin{equation}
\epsilon\equiv\frac{\lambda^2}{1-\lambda^2}=0.05,
\end{equation}
and ${\cal C}'$ and $d'e^{i\theta'}$ are the $B_s^0\to K^+K^-$ counterparts of the
$B^0_d\to\pi^+\pi^-$ parameters ${\cal C}$ and $de^{i\theta}$, respectively. 
If we apply the $U$-spin symmetry, we obtain the following relations \cite{RF-BsKK}:
\begin{equation}\label{U-spin-rel-1}
d'=d, \quad \theta'=\theta.
\end{equation}
As was also pointed out in Ref.~\cite{RF-BsKK}, these relations are not affected
by factorizable $U$-spin-breaking corrections, i.e.\ the relevant form factors and
decay constants cancel. This feature holds also for chirally enhanced contributions
to the transition amplitudes.

Since the CP asymmetries of the $B^0_s\to K^+K^-$ decay have not yet been
measured, we have to use the CP-averaged branching ratio of this mode, which 
also provides valuable information. For the determination of $\gamma$, it is useful 
to introduce the quantity
\begin{equation}\label{K-det}
K=\frac{1}{\epsilon}\,\left|\frac{{\cal C}}{{\cal C}'}\right|^2
\left[\frac{M_{B_s}}{M_{B_d}}\,\frac{\Phi(M_\pi/M_{B_d},M_\pi/M_{B_d})}{
\Phi(M_K/M_{B_s},M_K/M_{B_s})}\,\frac{\tau_{B_d}}{\tau_{B_s}}\right]
\left[\frac{\mbox{BR}(B_s\to K^+K^-)}{\mbox{BR}(B_d\to\pi^+\pi^-)}\right],
\end{equation}
where 
\begin{equation}
\Phi(x,y)\equiv\sqrt{\left[1-(x+y)^2\right]\left[1-(x-y)^2\right]}
\end{equation}
is the well-known $B\to PP$ phase-space function, and
the $\tau_{B_{d,s}}$ are the $B_{d,s}$ lifetimes. Applying the relations
in (\ref{U-spin-rel-1}), we arrive at
\begin{equation}\label{K-expr}
K=\frac{1}{\epsilon^2}\left[\frac{\epsilon^2+2\epsilon d\cos\theta\cos\gamma+
d^2}{1-2d\cos\theta\cos\gamma+d^2}\right].
\end{equation}
If we combine $K$ with ${\cal A}_{\rm CP}^{\rm mix}(B_d\to \pi^+\pi^-)$, which depend
both on $d\cos\theta$, we can fix another contour in the $\gamma$--$d$ plane with 
the help of the formulae given in Ref.~\cite{RF-BsKK}. 

In order to determine $K$ from the CP-averaged branching ratios, the $U$-spin-breaking 
corrections to the ratio $|{\cal C}'/{\cal C}|$, which equals 1 in the strict $U$-spin limit, 
have to be determined. In contrast to the $U$-spin relations in (\ref{U-spin-rel-1}), 
$|{\cal C}'/{\cal C}|$ involves hadronic form factors in the factorization approximation:
\begin{equation}\label{CpC-fact}
\left|\frac{{\cal C}'}{{\cal C}}\right|_{\rm fact}=
\frac{f_K}{f_\pi}\frac{F_{B_sK}(M_K^2;0^+)}{F_{B_d\pi}(M_\pi^2;0^+)}
\left(\frac{M_{B_s}^2-M_K^2}{M_{B_d}^2-M_\pi^2}\right),
\end{equation}
where $f_K$ and $f_\pi$ denote the kaon and pion decay constants, and 
$F_{B_sK}(M_K^2;0^+)$ and $F_{B_d\pi}(M_\pi^2;0^+)$ parametrize the 
hadronic quark-current matrix elements 
$\langle K^-|(\bar b u)_{\rm V-A}|B^0_s\rangle$ and 
$\langle\pi^-|(\bar b u)_{\rm V-A}|B^0_d\rangle$, respectively \cite{BSW}.
These quantities were analyzed using QCD sum-rule techniques in detail 
in Ref.~\cite{KMM}, yielding
\begin{equation}\label{SR-1}
\left|\frac{{\cal C}'}{{\cal C}}\right|^{\rm QCDSR}_{\rm fact}=1.52^{+0.18}_{-0.14}.
\end{equation}
As we will see in Section~\ref{sec:BspiK}, we can actually determine this quantity 
with the help of the data for the $B_d\to\pi^\mp K^\pm$, $B_s\to \pi^\pm K^\mp$
system. Since the corresponding value agrees remarkably well with 
(\ref{SR-1}), large non-factorizable $U$-spin-breaking effects are 
disfavoured, which gives us further confidence in applying (\ref{U-spin-rel-1}).

\boldmath
\subsection{Extraction of $\gamma$ and Hadronic Parameters}
\unboldmath
If we use (\ref{BR-Bdpipi}) and (\ref{BR-BsKK}) with (\ref{SR-1}) and add the errors 
in quadrature, we obtain
\begin{equation}\label{K-exp}
K=41.03\pm 10.27.
\end{equation}
In Fig.~\ref{fig:BdpipiBsKK-cont1}, we have also included the contour following from 
the central values of $K$ and ${\cal A}_{\rm CP}^{\rm mix}(B_d\to \pi^+\pi^-)$. We 
see that the intersections with the 
${\cal A}_{\rm CP}^{\rm dir}(B_d\to \pi^+\pi^-)$--${\cal A}_{\rm CP}^{\rm mix}
(B_d\to \pi^+\pi^-)$ contour following from the BaBar data give a twofold solution 
for $\gamma$ around $41^\circ$ and $67^\circ$, whereas we obtain no intersection
with the corresponding Belle curve. Consequently, the measured $B_s\to K^+K^-$
branching ratio disfavours the Belle result for the direct CP violation in
$B^0_d\to\pi^+\pi^-$. A similar observation was also made in Ref.~\cite{FRS},
using, however, a different avenue. For the following analysis, we will therefore only 
use the BaBar measurement of ${\cal A}_{\rm CP}^{\rm dir}(B_d\to \pi^+\pi^-)$, which 
covers also the prediction for this asymmetry made in Ref.~\cite{FRS} within the 
uncertainties. 

\begin{figure}
\centerline{
\begin{tabular}{cc}
 \includegraphics[width=7.5truecm]{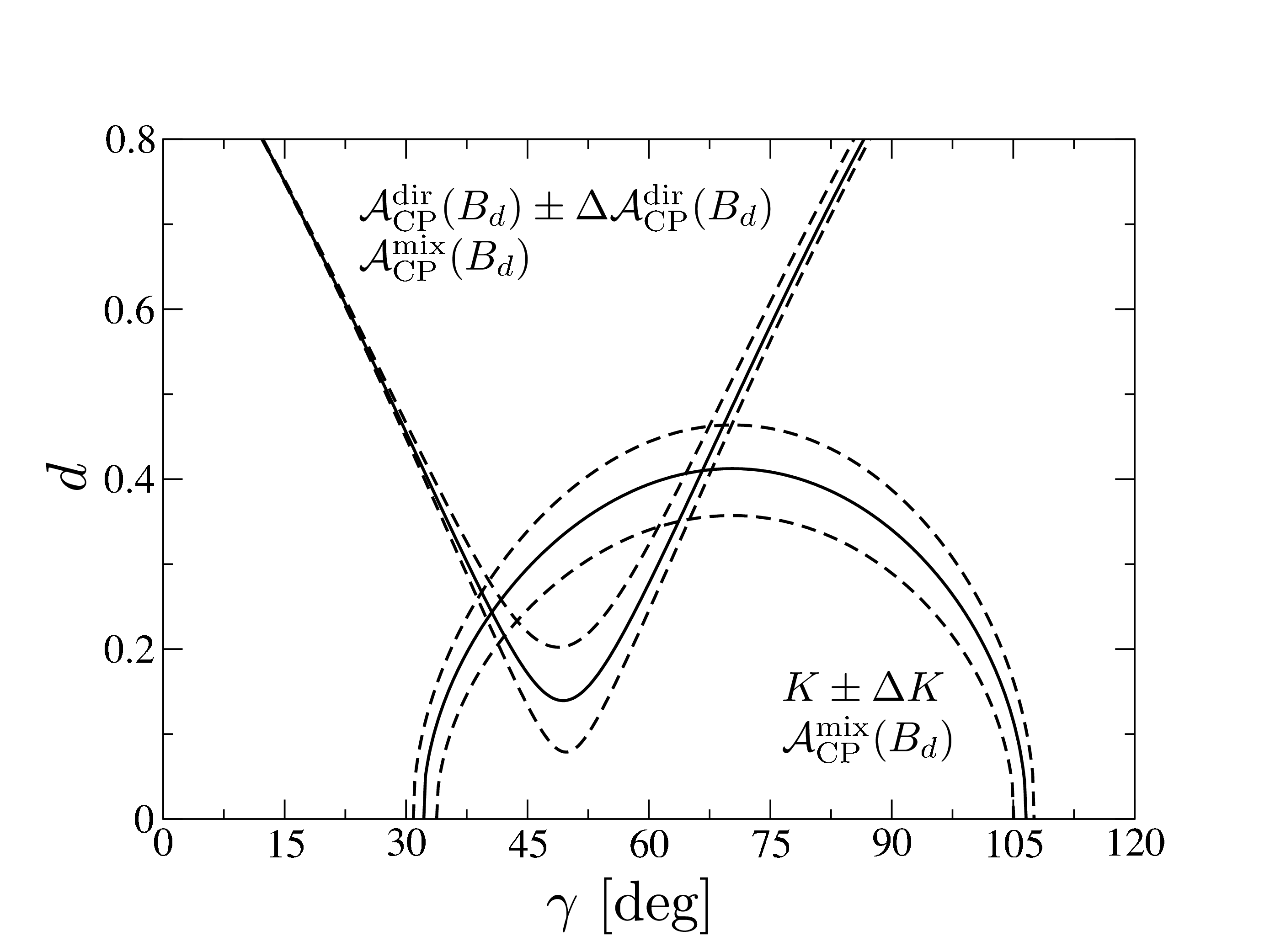} &
 \includegraphics[width=7.5truecm]{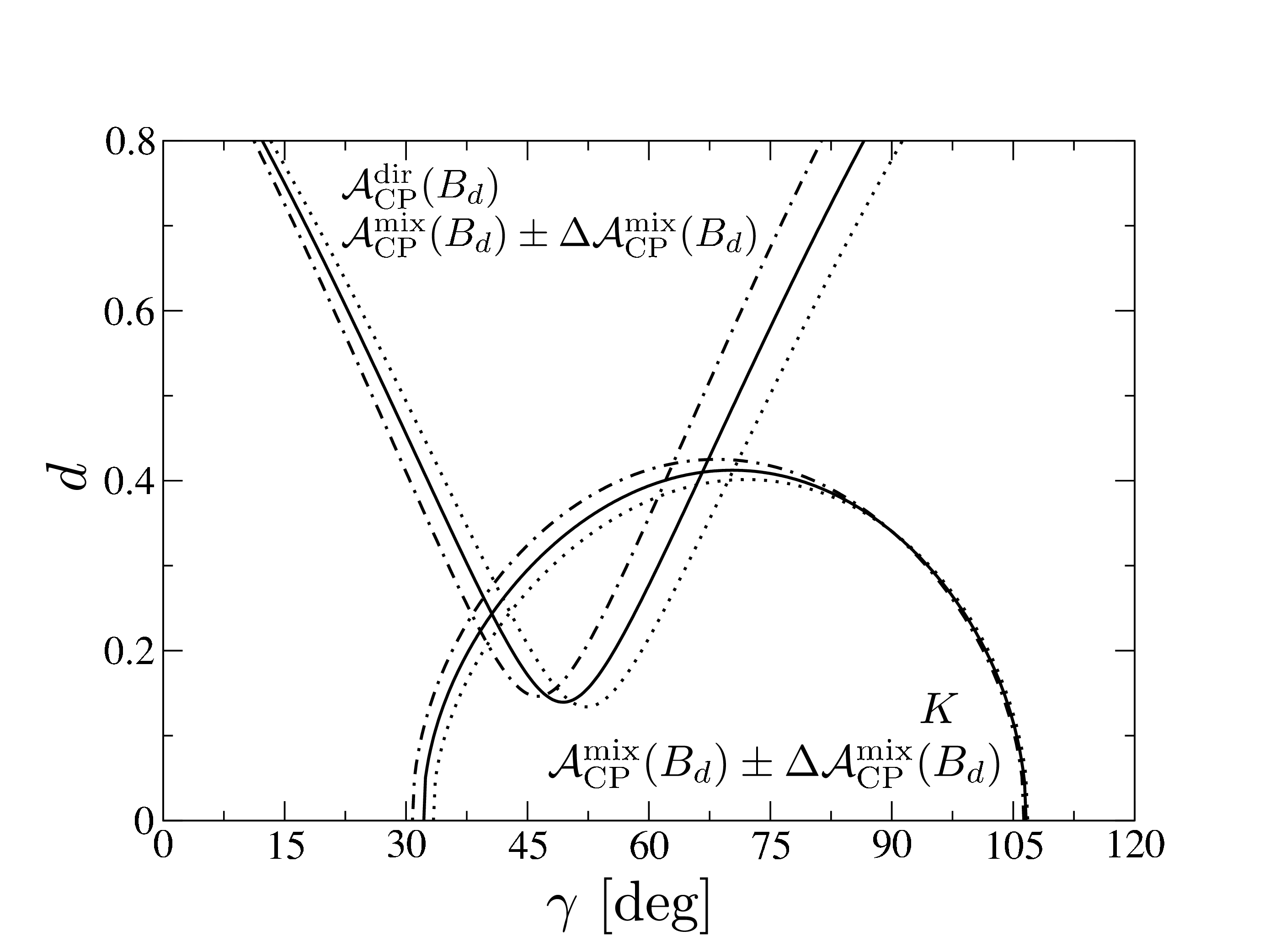}
 \end{tabular}}
 \vspace*{-0.3truecm}
\caption{Contours in the $\gamma$--$d$ plane fixed through the CP 
asymmetries of $B^0_d\to\pi^+\pi^-$ for the BaBar result of direct
CP violation and the quantity $K$: the left panel shows the 1\,$\sigma$ ranges 
of $K$ (upper and lower curves correspond to $K=51.30$ and $30.76$, respectively) 
and ${\cal A}_{\rm CP}^{\rm dir}(B_d\to \pi^+\pi^-)$ (upper and lower curves 
correspond to ${\cal A}_{\rm CP}^{\rm dir}=-0.30$ and $-0.12$, respectively), 
whereas the right panel shows the 1\,$\sigma$ range of  
${\cal A}_{\rm CP}^{\rm mix}(B_d\to \pi^+\pi^-)$ (dot-dashed and dotted curves 
correspond to ${\cal A}_{\rm CP}^{\rm mix}=0.69$ and $0.53$, 
respectively).}\label{fig:BdpipiBsKK-cont2}
\end{figure}

In Fig.~\ref{fig:BdpipiBsKK-cont2}, we show impact of the uncertainties of $K$ 
and the CP asymmetries of $B^0_d\to\pi^+\pi^-$. We obtain the following 
numerical results:
\begin{equation}\label{BsKKBdpipi-1}
\begin{array}{rclll}
\gamma&=&(40.6^{+1.6+1.1+2.3}_{-1.3-0.6-2.4})^\circ
& = &(40.6^{+3.0}_{-2.8})^\circ,\\
d&=&0.243^{+0.024+0.015+0.002}_{-0.028-0.008-0.001} & = &
0.243^{+0.028}_{-0.029},\\
\theta&=&(29.2^{+5.5+14.2+1.7}_{-3.5-12.8-1.3})^\circ & = &
(29.2^{+15.3}_{-13.3})^\circ,
\end{array}
\end{equation}
\begin{equation}\label{BsKKBdpipi-2}
\begin{array}{rclll}
\gamma & = &(66.6^{+2.6+1.1+3.2}_{-2.9-2.0-3.6})^\circ & = &
(66.6^{+4.3}_{-5.0})^\circ,\\
 d & = & 0.410^{+0.053+0.001+0.010}_{-0.060-0.003-0.009} & = &
 0.410^{+0.054}_{-0.061},\\
 \theta & = & (155.9^{+2.5+10.8+0.8}_{-3.8-2.1-1.2})^\circ & = &
(155.9^{+11.1}_{-4.5})^\circ,
\end{array}
\end{equation}
Here we show the errors arising from $K$, ${\cal A}_{\rm CP}^{\rm dir}
(B_d\to \pi^+\pi^-)$ and ${\cal A}_{\rm CP}^{\rm mix}(B_d\to \pi^+\pi^-)$,
and have finally added them in quadrature.

\boldmath
\subsection{Impact of $U$-Spin-Breaking Effects}\label{ssec:U-break}
\begin{figure}
\centerline{
 \includegraphics[width=7.5truecm]{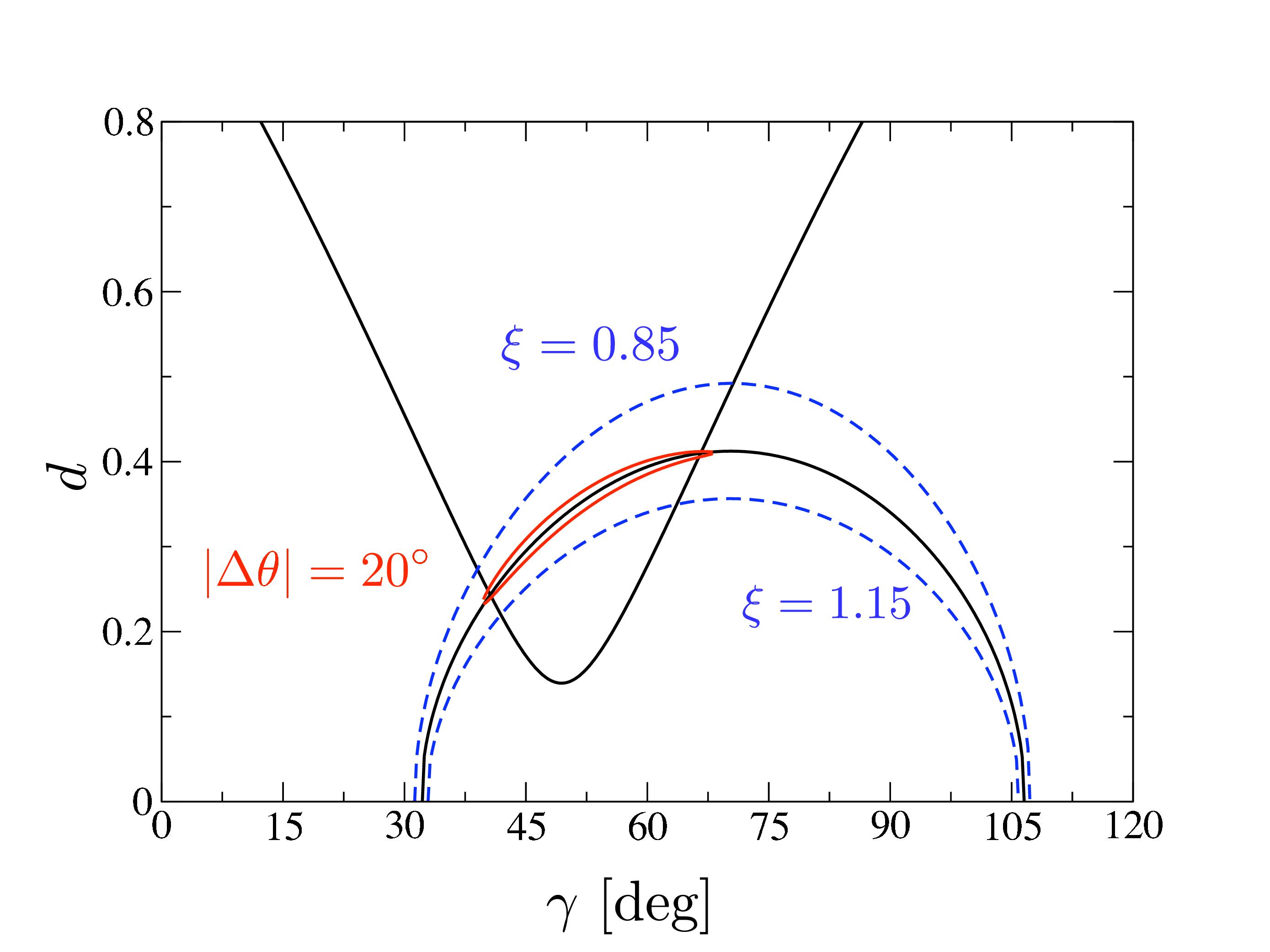}}
 \vspace*{-0.3truecm}
\caption{Illustration of the impact of $U$-spin-breaking corrections in
the $\gamma$--$d$ plane.}\label{fig:gam-d-break}
\end{figure}

\unboldmath
Let us now explore the impact of non-factorizable $U$-spin-breaking corrections
to (\ref{U-spin-rel-1}) by introducing the following parameters \cite{RF-00,FlMa-2}:
\begin{equation}
\xi\equiv d'/d, \quad \Delta\theta\equiv\theta'-\theta.
\end{equation}
The expression for $K$ in (\ref{K-expr}) is then modified as
\begin{equation}\label{K-break}
K=\frac{1}{\epsilon^2}\left[\frac{\epsilon^2+2\epsilon \xi d\cos(\theta+\Delta\theta)
\cos\gamma+\xi^2d^2}{1-2d\cos\theta\cos\gamma+d^2}\right].
\end{equation}
Since the numerator is governed by the $\xi^2d^2$ term, the dominant 
$U$-spin-breaking effects are described by $\xi$, whereas $\Delta\theta$ 
plays a very minor r\^ole, as was also noted in Refs.~\cite{RF-00,FlMa-2}.
This behaviour can nicely be seen in Fig.~\ref{fig:gam-d-break}, where we
have considered $\xi=1\pm0.15$ and $\Delta\theta=\pm20^\circ$. In view of
the comments given above, these parameters describe generous 
$U$-spin-breaking effects. Their impact on the numerical solutions in
(\ref{BsKKBdpipi-1}) and (\ref{BsKKBdpipi-2}) is given as follows:
\begin{equation}\label{BsKKBdpipi-1-U}
\begin{array}{rcl}
\gamma&=&(40.6^{+3.0+1.3+0.2}_{-2.8-1.6-0.3})^\circ,\\
d&=&0.243^{+0.028+0.030+0.006}_{-0.029-0.023-0.003},\\
\theta&=&(29.2^{+15.3+4.5+0.5}_{-13.3-4.3-0.8})^\circ,
\end{array}
\end{equation}
\begin{equation}\label{BsKKBdpipi-2-U}
\begin{array}{rcl}
\gamma & = & (66.6^{+4.3+4.0+0.1}_{-5.0-3.0-0.2})^\circ,\\
 d & = & 0.410^{+0.054+0.082+0.002}_{-0.061-0.060-0.001},\\
 \theta & = & (155.9^{+11.1+3.6+0.1}_{-4.5-3.8-0.3})^\circ,
\end{array}
\end{equation}
where the second and third errors refer to $\xi$ and $\Delta\theta$, respectively. 
Interestingly, $\gamma$ is only moderately affected by these effects, which 
do not exceed the current experimental uncertainties for the parameter 
ranges considered above. Performing measurements of  CP violation in 
$B_s\to K^+K^-$, which will be possible with impressive accuracy at the LHCb 
experiment \cite{LHCb-Bhh}, the use of the $U$-spin symmetry can be minimized in 
the extraction of $\gamma$, and internal consistency checks become available.
Before turning to these asymmetries, let us first discuss the discrete 
ambiguities affecting the extraction of $\gamma$.

\subsection{Discrete Ambiguities}\label{ssec:ambig}
So far, we have restricted the discussion to the range of
$0^\circ\leq\gamma\leq 180^\circ$, which follows from the SM interpretation
of the measurement of $\varepsilon_K$, which describes the indirect CP violation 
in the neutral kaon system \cite{UTfit,CKMfitter}. However, if we allow for 
new physics (NP), we have to consider the whole range of $\gamma$. As can
be seen by having a closer look at the expressions given in (\ref{ACP-dir-d}), 
(\ref{ACP-mix-d}) and (\ref{K-expr}), for each of the two solutions listed in
(\ref{BsKKBdpipi-1-U}) and (\ref{BsKKBdpipi-2-U}), we obtain an additional one
through the following transformation:
\begin{equation}\label{ambig-1}
\gamma\to\gamma-180^\circ, \quad d\to d, \quad \theta\to\theta-180^\circ,
\end{equation}
i.e.\ we have to deal with a fourfold discrete ambiguity, which has to be
resolved for the search of NP. 

To this end, let us first have a look at $\cos\theta$ for (\ref{BsKKBdpipi-1-U}) and 
(\ref{BsKKBdpipi-2-U}), given by 
\begin{equation}
\cos\theta=+0.873^{+0.092}_{-0.168} \quad\mbox{and}\quad 
\cos\theta=-0.913^{+0.047}_{-0.064}, 
\end{equation}
respectively, where we have added all errors in quadrature. 
Although non-factorizable effects
have a significant impact on $\theta$, we do {\it not} expect that they will
change the sign of the cosine of this strong phase, which is {\it negative}
in the notation used above. Consequently,  (\ref{BsKKBdpipi-1-U}) can be 
excluded through this argument. As we will see in Subsection~\ref{ssec:ACPmix},
the future measurement of mixing-induced CP violation in $B_s\to K^+K^-$ should
allow us to rule out this solution in a {\it direct} way. Moreover, as will be discussed 
in Subsection~\ref{ssec:contact}, already the current data for the observables of
the $B_d\to\pi^\mp K^\pm$, $B^\pm\to\pi^\pm K$ system exclude
(\ref{BsKKBdpipi-1-U}) and its ``mirror'' solution around $\gamma=-139^\circ$ 
following from (\ref{ambig-1}), where the sign of $\cos\theta$ would be as in 
factorization. In the case of the remaining mirror solution of (\ref{BsKKBdpipi-2-U}) 
around $\gamma=-113^\circ$, the sign of $\cos\theta$ would be positive, 
i.e.\ opposite to our expectation, so that it can be ruled out as well. 

Consequently, we are finally left with the numbers in (\ref{BsKKBdpipi-2-U}). It is 
interesting to note that the corresponding value of $\gamma$ in is in excellent 
agreement with the SM fits of the UT obtained by the UTfit and CKMfitter collaborations 
\cite{UTfit,CKMfitter}, yielding $\gamma=(64.6 \pm 4.2)^\circ$ and 
$\gamma=(59.0^{+9.2}_{-3.7})^\circ$, respectively.

\boldmath
\subsection{CP Violation in $B_s\to K^+K^-$}\label{ssec:ACPmix}
\unboldmath
Using the expression for the $B^0_s\to K^+K^-$ decay amplitude in 
(\ref{BsKK-ampl}), the observables entering the CP-violating rate asymmetry
in (\ref{ACP-t}) take the following form:
\begin{eqnarray}
{\cal A}_{\rm CP}^{\rm dir}(B_s\to K^+K^-) & = &
\frac{2\epsilon d'\sin\theta'\sin\gamma}{d'^2+
2\epsilon d'\cos\theta'\cos\gamma+\epsilon^2},\label{Adir-BsKK}\\
{\cal A}_{\rm CP}^{\rm mix}(B_s\to K^+K^-)&=&+\left[
\frac{d'^2\sin\phi_s + 2\epsilon d' \cos\theta' \sin(\phi_s+\gamma)
+ \epsilon^2\sin(\phi_s+2\gamma)}{d'^2+2\epsilon d'\cos\theta'\cos\gamma+
\epsilon^2}\right],\label{Amix-BsKK}\\
{\cal A}_{\Delta\Gamma}(B_s\to K^+K^-)&=&
-\left[\frac{d'^2\cos\phi_s+2\epsilon d'\cos\theta'\cos(\phi_s+\gamma)
+\epsilon^2\cos(\phi_s+2\gamma)}{d'^2+2\epsilon d'\cos\theta'\cos\gamma+
\epsilon^2d'^2}\right],\qquad\mbox{}\label{ADG-BsKK}
\end{eqnarray}
where $\phi_s$ is the CP-violating $B^0_s$--$\bar B^0_s$ mixing phase;
in the SM, it is given in terms of the Wolfenstein parameters by 
$\phi_s=-2\lambda^2\eta$, and takes a tiny value of $\phi_s|_{\rm SM}\approx -2 ^\circ$.
If we consider this SM case for the solution of (\ref{BsKKBdpipi-2})
and use the $U$-spin relations in (\ref{U-spin-rel-1}), we arrive at the 
following predictions:
\begin{equation}\label{ACP-BsKK}
\begin{array}{rclll}
{\cal A}_{\rm CP}^{\rm dir}(B_s\to K^+K^-)&=&
+0.101^{+0.034+0.043+0.000}_{-0.020-0.043-0.000}&=&+0.101^{+0.055}_{-0.047},\\
{\cal A}_{\rm CP}^{\rm mix}(B_s\to K^+K^-)&=&
-0.246^{+0.018+0.029+0.012}_{-0.023-0.017-0.010}&=&-0.246^{+0.036}_{-0.030},\\
{\cal A}_{\Delta\Gamma}(B_s\to K^+K^-)&=&
-0.964^{+0.010+0.001+0.003}_{-0.006-0.002-0.003}&=&-0.964^{+0.011}_{-0.007},
\end{array}
\end{equation}
where the treatment and notation of the errors is as in (\ref{BsKKBdpipi-1}) 
and (\ref{BsKKBdpipi-2}), i.e. refers to the uncertainties of $K$, 
${\cal A}_{\rm CP}^{\rm dir}(B_d\to\pi^+\pi^-)$ and
${\cal A}_{\rm CP}^{\rm mix}(B_d\to\pi^+\pi^-)$. The interesting
feature that the error of the direct CP asymmetry is independent of that of 
${\cal A}_{\rm CP}^{\rm mix}(B_d\to\pi^+\pi^-)$ is due to the following 
$U$-spin relation \cite{RF-BsKK}:
\begin{equation}\label{ACP-BR-rel1}
{\cal A}_{\rm CP}^{\rm dir}(B_s\to K^+K^-)=-\frac{1}{\epsilon K} 
{\cal A}_{\rm CP}^{\rm dir}(B_d\to \pi^+\pi^-),
\end{equation}
and provides a nice numerical test. Moreover, all observables satisfy the relation
\begin{equation}
\left[{\cal A}_{\rm CP}^{\rm dir}(B_s\to K^+K^-) \right]^2+
\left[ {\cal A}_{\rm CP}^{\rm mix}(B_s\to K^+K^-)\right]^2+
\left[ {\cal A}_{\Delta\Gamma}(B_s\to K^+K^-)\right]^2=1.
\end{equation}
The impact of the $U$-spin-breaking corrections discussed in 
Subsection~\ref{ssec:U-break} is given as follows:
\begin{equation}\label{ACP-BsKK-break}
\begin{array}{rcl}
{\cal A}_{\rm CP}^{\rm dir}(B_s\to K^+K^-)&=&+0.101^{+0.055+0.015+0.067}_{-0.047
-0.015-0.083},\\
{\cal A}_{\rm CP}^{\rm mix}(B_s\to K^+K^-)&=&-0.246^{+0.036+0.008+
0.051}_{-0.030-0.007-0.023},\\
{\cal A}_{\Delta\Gamma}(B_s\to K^+K^-)&=&-0.964^{+0.011+0.000+0.001}_{-0.007-
0.000-0.002},
\end{array}
\end{equation}
where the second and third errors refer to $\xi=1\pm0.15$ and
$\Delta\theta=\pm20^\circ$, respectively, as in (\ref{BsKKBdpipi-1-U})
and  (\ref{BsKKBdpipi-2-U}). Whereas ${\cal A}_{\rm CP}^{\rm mix}(B_s\to K^+K^-)$ 
and ${\cal A}_{\Delta\Gamma}(B_s\to K^+K^-)$ are pretty stable with respect
to the $U$-spin-breaking effects, the direct CP asymmetry is significantly affected
by $\Delta\theta$. As we will discuss in Subsection~\ref{ssec:contact}, the
measurement of the direct CP violation in $B_d\to\pi^\mp K^\pm$
strongly disfavours such effects.

\begin{figure}
\centerline{
 \includegraphics[width=7.5truecm]{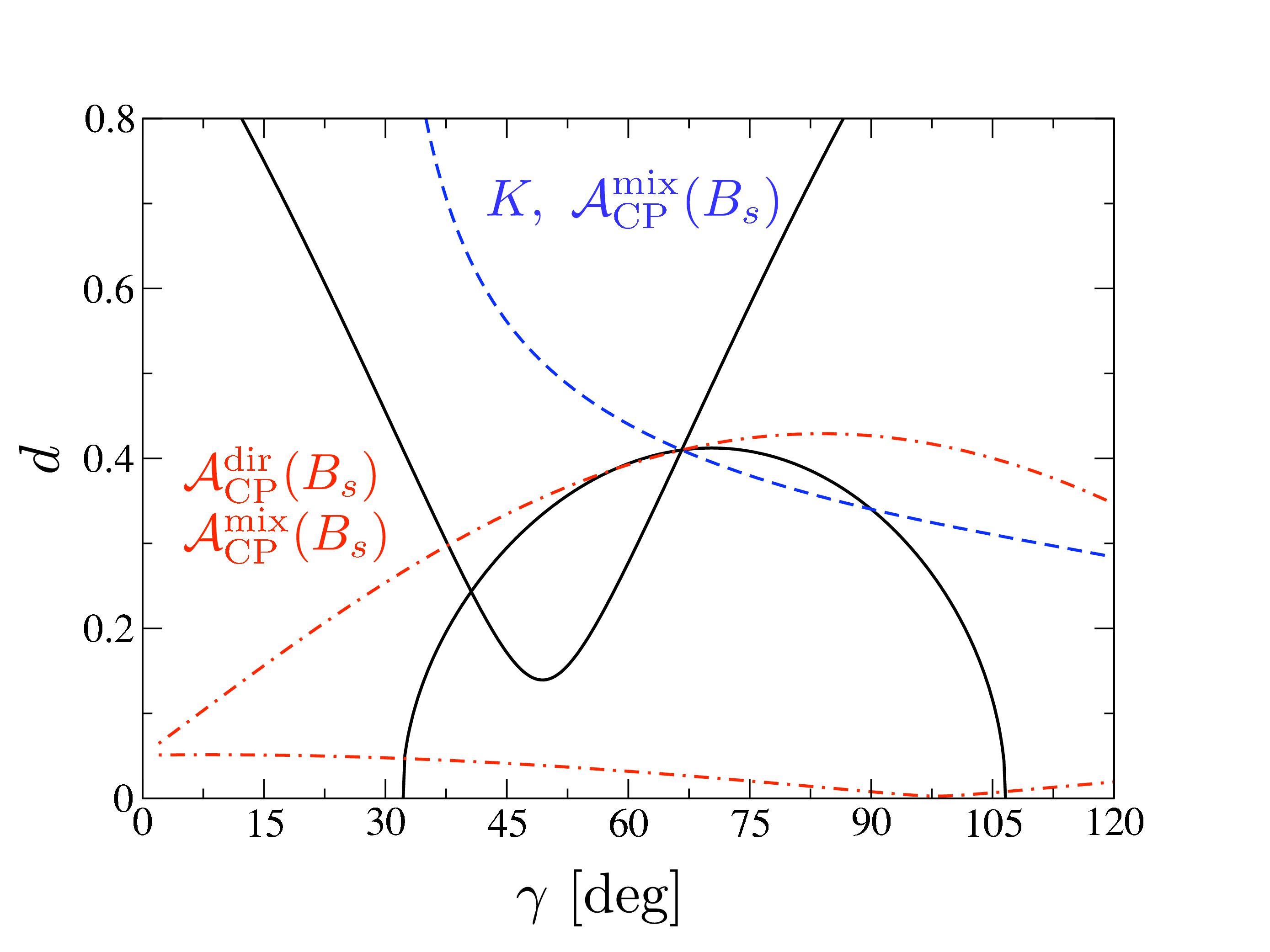}}
 \vspace*{-0.3truecm}
\caption{Illustration of the impact of the measurement of the CP-violating
observables of the $B^0_s\to K^+K^-$ decay on the situation in the 
$\gamma$--$d$ plane within the SM.}\label{fig:BdpipiBsKK-cont3}
\end{figure}

The next important step in the analysis of the $B_d\to\pi^+\pi^-$, $B_s\to K^+K^-$
system is the measurement of the mixing-induced CP violation in $B^0_s\to K^+K^-$.
Applying the formulae given in Ref.~\cite{RF-BsKK}, this observable can
be combined with $K$ to fix another contour in the $\gamma$--$d$ plane.
In Fig.~\ref{fig:BdpipiBsKK-cont3}, we illustrate the corresponding situation 
for the central numerical values given above, and observe that the measurement 
of ${\cal A}_{\rm CP}^{\rm mix}(B_s\to K^+K^-)$ will in fact allow us to resolve the
twofold ambiguity in the extraction of the UT angle $\gamma$, as we 
noted in Subsection~\ref{ssec:ambig}.

Finally, if also the direct CP asymmetry ${\cal A}_{\rm CP}^{\rm dir}(B_s\to K^+K^-)$
is measured, we can combine it with ${\cal A}_{\rm CP}^{\rm mix}(B_s\to K^+K^-)$
to calculate $d'$ as a function of $\gamma$ for a given value of the mixing phase 
$\phi_s$ \cite{RF-BsKK}. It should be emphasized that this contour is -- in contrast to 
those involving $K$ -- {\it theoretically clean}, in analogy to the $\gamma$--$d$ curve following from the CP-violating $B_d\to\pi^+\pi^-$ observables.  Using the first of 
the $U$-spin relations in (\ref{U-spin-rel-1}), we can then extract $\gamma$ and 
$d$, where the information provided by $K$ allows us to resolve the discrete 
ambiguity. Since the strong phases $\theta$ and $\theta'$ can be determined as 
well, we may actually perform a test of the second $U$-spin relation in 
(\ref{U-spin-rel-1}). Moreover, the impact of $U$-spin-breaking corrections to 
$d'=d$ corresponds to a relative shift of the $B_d\to\pi^+\pi^-$ and $B_s\to K^+K^-$ 
contours; the situation for the extraction of $\gamma$ in 
Fig.~\ref{fig:BdpipiBsKK-cont3} would actually be very stable in this respect. 
This would be the most refined implementation of the $B_s\to K^+K^-$,
$B_d\to\pi^+\pi^-$ strategy for the extraction of $\gamma$. For recent LHCb 
studies, which look very promising, see Ref.~\cite{LHCb-Bhh}.

The last observable that is provided by $B^0_s\to K^+K^-$ is 
${\cal A}_{\Delta\Gamma}(B_s\to K^+K^-)$, which enters the 
following ``untagged" rate \cite{DF}:
\begin{eqnarray}
\lefteqn{\langle \Gamma(B_s(t)\to K^+K^-)\rangle  \equiv  
\Gamma(B^0_s(t)\to K^+K^-)+ \Gamma(\bar B^0_s(t)\to K^+K^-)}\nonumber\\
& & \propto  e^{-\Gamma_st}\left[e^{+\Delta\Gamma_s t/2}R_{\rm L}(B_s\to K^+K^-)
+e^{-\Delta\Gamma_s t/2}R_{\rm H}(B_s\to K^+K^-)\right],\label{untagged}
\end{eqnarray}
where 
\begin{equation}\label{DG-def}
\Gamma_s\equiv\frac{\Gamma_{\rm H}^{(s)}+\Gamma_{\rm L}^{(s)}}{2}, \quad
\Delta\Gamma_s\equiv\Gamma_{\rm H}^{(s)}-\Gamma_{\rm L}^{(s)}
\end{equation}
depend on the decay widths $\Gamma_{\rm H}^{(s)}$ and $\Gamma_{\rm L}^{(s)}$ 
of the ``heavy" and ``light" mass eigenstates of the $B_s$ system, respectively, and
\begin{eqnarray}
R_{\rm L}(B_s\to K^+K^-)&\equiv&
1-{\cal A}_{\Delta\Gamma}(B_s\to K^+K^-)=1.964^{+0.007}_{-0.011},\\
R_{\rm H}(B_s\to K^+K^-)&\equiv&1+{\cal A}_{\Delta\Gamma}(B_s\to K^+K^-)
=0.036^{+0.011}_{-0.007};
\end{eqnarray}
the numerical values correspond to the SM prediction in
(\ref{ACP-BsKK-break}). Concerning a practical measurement of (\ref{untagged}), 
most data come from short times with $\Delta\Gamma_st\ll1$:
\begin{equation}\label{BsKK-UT}
\langle \Gamma(B_s(t)\to K^+K^-)\rangle\propto
e^{-\Gamma_st}\left[1-{\cal A}_{\Delta\Gamma}(B_s\to K^+K^-)
\left(\frac{\Delta\Gamma_st}{2}\right)+{\cal O}((\Delta\Gamma_st)^2)\right].
\end{equation}
Moreover, if the two-exponential form of (\ref{untagged}) is fitted to a single 
exponential, the corresponding decay width satisfies the following relation \cite{DFN}:
\begin{equation}\label{Gamma-fit}
\Gamma_{K^+K^-}=\Gamma_s+{\cal A}_{\Delta\Gamma}(B_s\to K^+K^-)
\frac{\Delta\Gamma_s}{2}+{\cal O}((\Delta\Gamma_s)^2/\Gamma_s).
\end{equation}
First studies along these lines were recently performed by 
the CDF collaboration \cite{CDF-DGBsKK}, yielding
$\tau(B_s\to K^+K^-)=1/\Gamma_{K^+K^-} = (1.53 \pm 0.18 \pm 0.02)\,\mbox{ps}$.
Using flavour-specific $B_s$ decays, a similar analysis allows the extraction 
of $\Gamma_s$ up to corrections of ${\cal O}((\Delta\Gamma_s/\Gamma_s)^2)$
\cite{DFN}. With the help of the analysis discussed above, which allows the
calculation of ${\cal A}_{\Delta\Gamma}(B_s\to K^+K^-)$, the width difference
$\Delta\Gamma_s$ can then be extracted.

\boldmath
\subsection{Impact of New Physics}
\unboldmath
Because of the impressive agreement of the value of $\gamma$ that we 
extracted from the $B_d\to\pi^+\pi^-$, $B_s\to K^+K^-$ data with the fits 
of the UT and the overall consistency with the SM (see also 
Section~\ref{sec:BspiK}), dramatic NP contributions to the corresponding 
decay amplitudes are already excluded, although the experimental picture 
has still to be improved considerably. In particular, accurate measurements 
of $\gamma$ through pure tree-level decays are not yet available, but will be 
performed at LHCb \cite{LHCb}; imporant examples are $B_s\to D_s^\pm K^\mp$ 
and $B_d\to D^\pm\pi^\mp$ decays, where the $U$-spin symmetry provides 
again a useful tool \cite{RF-BsDK}.

Similar conclusions about NP effects in $B\to\pi\pi, \pi K$ modes were 
drawn in Refs.~\cite{FRS,BFRS}. The corresponding $B$-factory data may 
indicate a modified EW penguin sector with a large CP-violating NP phase
through the results for mixing-induced CP violation in $B^0_d\to\pi^0 K_{\rm S}$,
thereby complementing the pattern of such CP asymmetries observed in other 
$b\to s$ penguin modes, where the $B^0_d\to\phi K_{\rm S}$ channel is an 
outstanding example. Since EW penguin topologies contribute to the 
$B_s\to K^+K^-$, $B_d\to\pi^+\pi^-$ (and the $B_d\to\pi^\mp K^\pm$, 
$B_s\to\pi^\pm K^\mp$) system in colour-suppressed form, they play there 
a minor r\^ole. Consequently, NP effects entering through the EW penguin 
sector could not be seen in the analysis discussed in this paper.

\begin{figure}
\centerline{
 \includegraphics[width=7.5truecm]{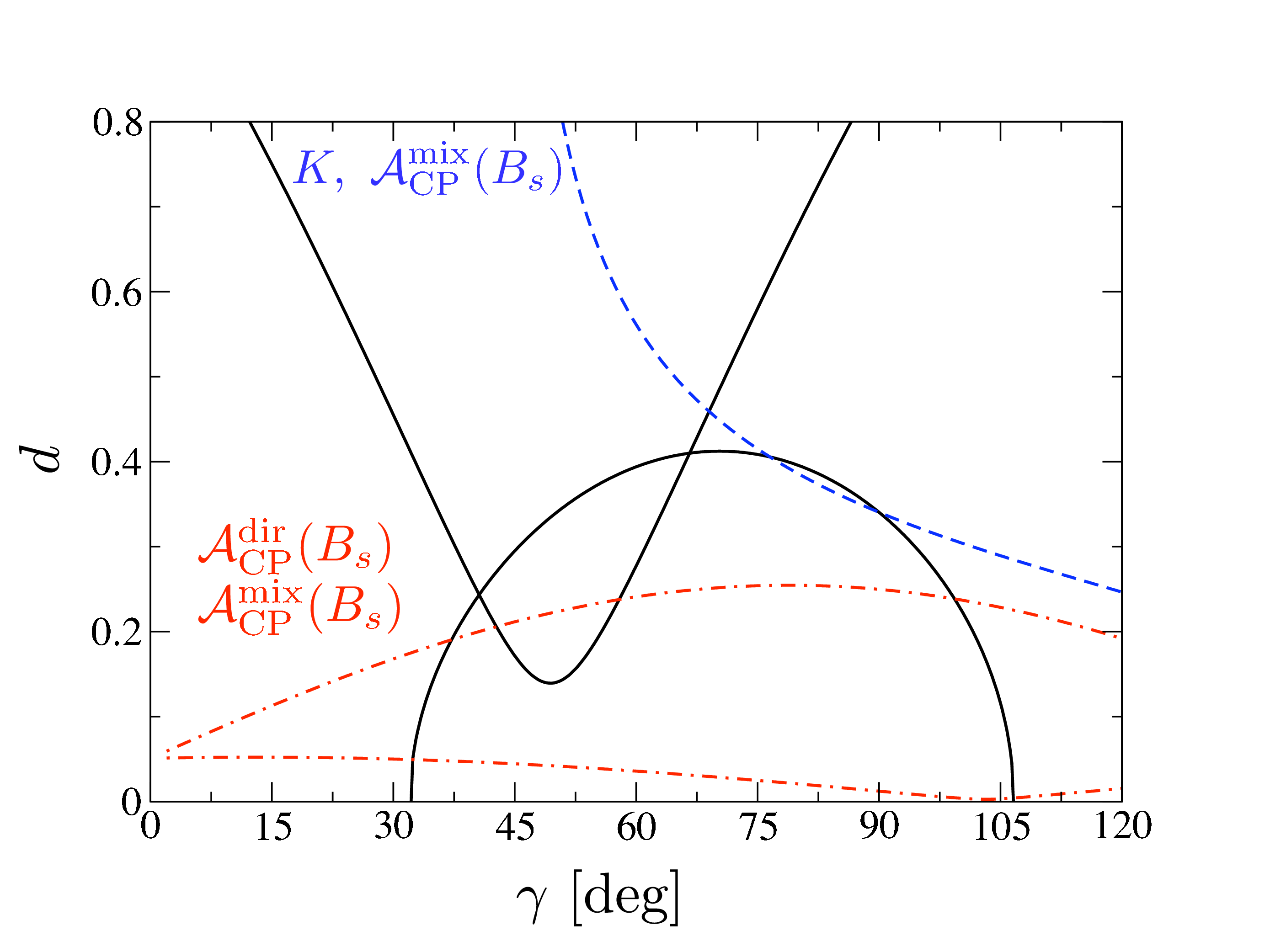}}
 \vspace*{-0.3truecm}
\caption{Illustration of the impact of CP-violating NP contributions to
$B^0_s$--$\bar B^0_s$ mixing leading to $\phi_s=-10^\circ$ on the
contours in the $\gamma$--$d$ plane.}\label{fig:Bs-NP-1}
\end{figure}

On the other hand, $B^0_s$--$\bar B^0_s$ mixing offers a nice avenue for
NP to manifest itself in $B^0_s\to K^+K^-$. The mass difference $\Delta M_s$
was recently measured at the Tevatron  \cite{D0-DMs,CDF-DMs}, with a 
value that is consistent with the SM expectation. On the other hand, this
result still allows for large CP-violating NP contributions to $B^0_s$--$\bar B^0_s$ 
mixing (see, for instance, Refs.~\cite{BF-06,DMspapers}). In this case, the
mixing phase $\phi_s$, which can be extracted through the  
time-dependent angular distribution of the 
$B^0_s\to J/\psi[\to\mu^+\mu^-]\phi[\to K^+K^-]$ decay products \cite{DDF,DFN},
would take a sizeable value. 
Interestingly, also the $B_s\to K^+K^-$, $B_d\to\pi^+\pi^-$ system allows us to 
search for NP effects of this kind. Assuming a value of 
$\phi_s=-10^\circ$, which corresponds 
to a simple ``translation" of the tension in the CKM fits between 
$(\sin2\beta)_{\psi K_{\rm S}}$ and the UT side 
$R_b\propto |V_{ub}/V_{cb}|$ \cite{BF-06}, we arrive at the situation 
illustrated in Fig.~\ref{fig:Bs-NP-1}. There we show the contours involving 
${\cal A}_{\rm CP}^{\rm mix}(B_s\to K^+K^-)$ that would arise if we 
assume the SM value of $\phi_s$. In this case, we would arrive at quite 
some discrepancy, in particular through the contour following from the 
CP-violating $B_s\to K^+K^-$ asymmetries. For larger values of $\phi_s$, 
the discrepancy would be even more pronounced. In this case, the measured
value of ${\cal A}_{\rm CP}^{\rm mix}(B_s\to K^+K^-)$ would also not lie on 
the SM surface in observable space that was calculated in Ref.~\cite{FM-space}.

\begin{figure}
\centerline{
 \includegraphics[width=7.5truecm]{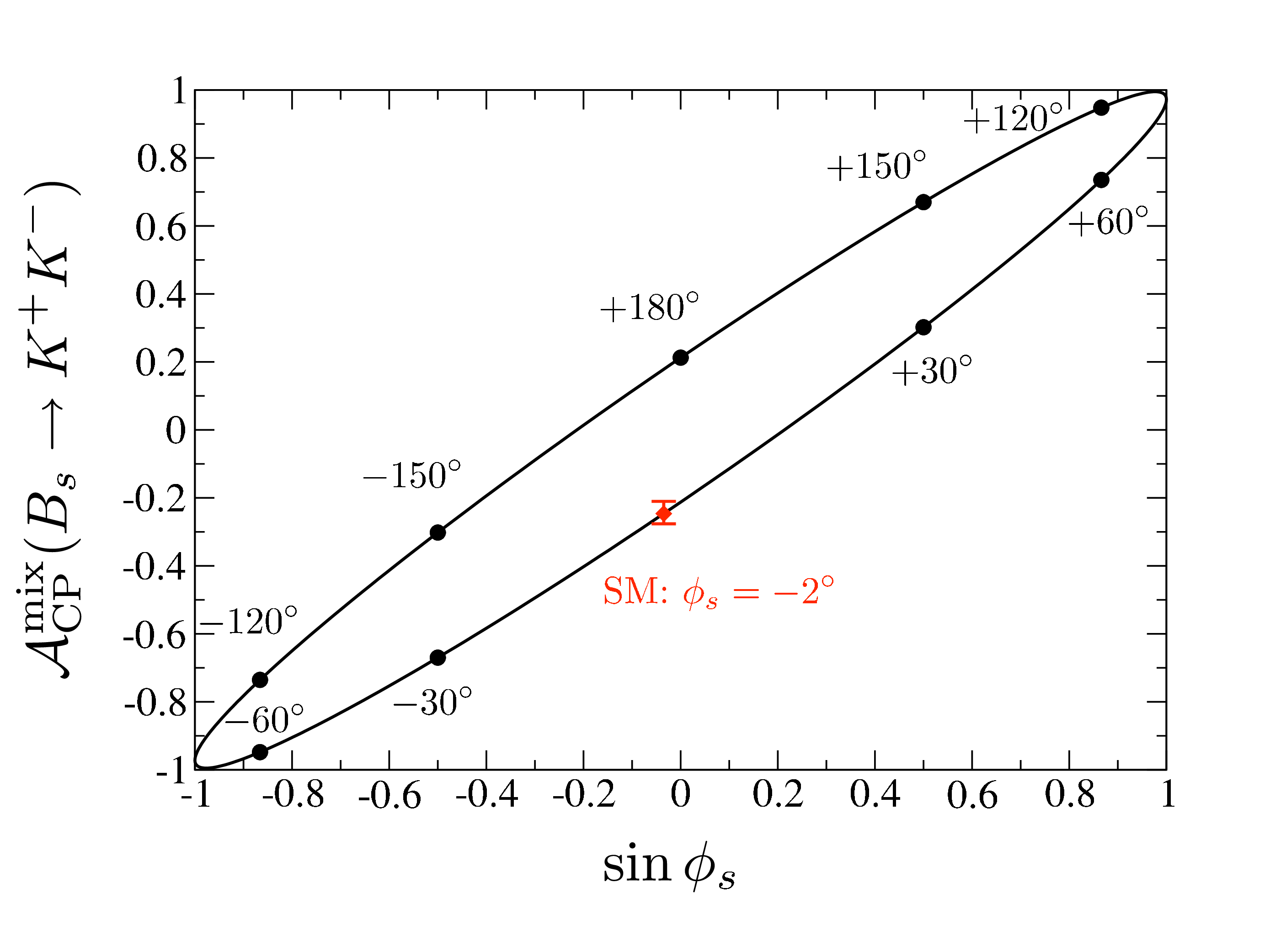}}
 \vspace*{-0.3truecm}
\caption{The correlation between $\sin\phi_s$, which can be determined 
through mixing-induced CP violation in $B^0_s\to J/\psi \phi$, and 
${\cal A}_{\rm CP}^{\rm mix}(B_s\to K^+K^-)$ for the central values of the
parameters in (\ref{BsKKBdpipi-2}); in the SM case, we show also the
error bar. Each point on the curve corresponds to a given value of $\phi_s$, 
as indicated by the numerical values.}\label{fig:Bs-NP-2}
\end{figure}

It is instructive to expand (\ref{Amix-BsKK}) and (\ref{ADG-BsKK}) in powers 
of $\epsilon/d'\sim0.1$, yielding
\begin{eqnarray}
{\cal A}_{\rm CP}^{\rm mix}(B_s\to K^+K^-)&=&+\sin\phi_s+2\left(\frac{\epsilon}{d'}\right)
\cos\theta'\sin\gamma\cos\phi_s+{\cal O}\left((\epsilon/d')^2\right),\\
{\cal A}_{\Delta\Gamma}(B_s\to K^+K^-)&=&-\cos\phi_s+2\left(\frac{\epsilon}{d'}\right)
\cos\theta'\sin\gamma\sin\phi_s+{\cal O}\left((\epsilon/d')^2\right),
\end{eqnarray}
where $2(\epsilon/d')\cos\theta'\sin\gamma\approx -0.2$. We observe two
interesting features:
\begin{itemize}
\item ${\cal A}_{\rm CP}^{\rm mix}(B_s\to K^+K^-)$ is strongly affected 
if $\phi_s$ moves away from $0$ thanks to the $\sin\phi_s$ term, and
offers also information on $\cos\phi_s$ through the hadronic piece.
\item ${\cal A}_{\Delta\Gamma}(B_s\to K^+K^-)$ deviates slowly from its SM
value around $-1$ as $\phi_s$ moves away from $0$, and the hadronic term 
is suppressed by $\sin\phi_s$ for small phases, which is the reason for the
remarkably small uncertainty of the SM prediction in Subsection~\ref{ssec:ACPmix}.
\end{itemize}
In Fig.~\ref{fig:Bs-NP-2}, we show the correlation between $\sin\phi_s$,
which is determined through the time-dependent angular analysis of the 
$B^0_s\to J/\psi [\to\mu^+\mu^-]\phi[\to K^+K^-]$ decay products \cite{DDF}, 
and the mixing-induced CP violation in $B^0_s\to K^+K^-$, which can be 
predicted with the help of the parameters in (\ref{BsKKBdpipi-2}). This figure 
shows nicely that the combination of both 
observables allows an {\it unambiguous} determination of $\phi_s$. In particular, 
we may also distinguish between the cases of $\phi_s=0^\circ$ and $180^\circ$, 
which is important for the search of NP. Recently, the D0 collaboration has
reported first results for the measurement of $\phi_s$ through 
an {\it untagged} $B^0_s\to J/\psi \phi$ analysis \cite{D0-phis}, which suffers
from a four-fold discrete ambiguity. The solution closest to the SM case reads as
\begin{equation}
\phi_s=-0.79\pm0.56\,\mbox{(stat.)} ^{+0.14}_{-0.01}\,\mbox{(syst.)}
=-(45\pm32^{+1}_{-8})^\circ,
\end{equation}
so that this quantity is still largely unconstrained. 

Let us finally come back to the untagged rate in (\ref{BsKK-UT}). In the presence 
of NP, $\Delta\Gamma_s$ is modified as follows \cite{grossman}:
\begin{equation}
\Delta\Gamma_s=\Delta\Gamma_s^{\rm SM}\cos\phi_s,
\end{equation}
where $\Delta\Gamma_s^{\rm SM}/\Gamma_s$ is negative for the definition in
(\ref{DG-def}), and calculated at the 15\% level~\cite{LN}. Consequently, NP 
effects can only reduce the value of $|\Delta\Gamma_s|$. If $\phi_s$ is
determined as described above, the calculation of 
${\cal A}_{\Delta\Gamma}(B_s\to K^+K^-)$ through the $B_s\to K^+K^-$, 
$B_d\to\pi^+\pi^-$ analysis allows the extraction of $\Delta\Gamma_s^{\rm SM}$ 
from the $\tau(B_s\to K^+K^-)$ lifetime, thereby complementing the extraction of 
this width difference through the $B_s\to J/\psi\phi$ angular analysis \cite{DDF}.

\boldmath
\section{The $B_d\to\pi^\mp K^\pm$, $B_s\to \pi^\pm K^\mp$ Strategy}\label{sec:BspiK}
\unboldmath
\setcounter{equation}{0}
\boldmath
\subsection{First Insights into $U$-Spin-Breaking Effects}
\unboldmath
Let us now discuss the $U$-spin-related decays $B^0_d\to\pi^-K^+$ and
$B^0_s\to\pi^+ K^-$ \cite{GR-U-spin}. If we use the unitarity of the CKM matrix, 
their decay amplitudes can be written as follows:
\begin{eqnarray}
A(B^0_d\to\pi^-K^+)&=& -P\left[1-r e^{i\delta}e^{i\gamma}\right],\label{B0pimKp-ampl}\\
A(B^0_s\to\pi^+K^-)&=& P_s\sqrt{\epsilon}\left[1+\frac{1}{\epsilon}r_s e^{i\delta_s}
e^{i\gamma}\right],
\end{eqnarray}
where $P_{(s)}$ and $r_{(s)} e^{i\delta_{(s)}}$ are CP-conserving hadronic parameters,
which describe penguin amplitudes and the ratio of trees to penguins, respectively. 
Using the $U$-spin flavour symmetry of strong interactions, we obtain -- in 
analogy to (\ref{U-spin-rel-1}) -- the following relations:
\begin{equation}\label{U-spin-rel-2}
r_s=r, \quad \delta_s=\delta.
\end{equation}
In the case of the relation between $|P_s|$ and $|P|$, factorizable $U$-spin-breaking
corrections arise, which are described by the following ratio of decay constants and 
form factors:
\begin{equation}\label{relP-Ps}
\left|\frac{P_s}{P} \right|_{\rm fact}=\frac{f_\pi}{f_K}
\frac{F_{B_sK}(M_\pi^2;0^+)}{F_{B_d\pi}(M_K^2;0^+)}
\left(\frac{M_{B_s}^2-M_K^2}{M_{B_d}^2-M_\pi^2}\right).
\end{equation}
Using the recent QCD sum-rule results of Ref.~\cite{KMM} yields
\begin{equation}\label{relP-Ps-QCDSR}
\left|\frac{P_s}{P} \right|_{\rm fact}^{\rm QCDSR}=1.02^{+0.11}_{-0.10}.
\end{equation}
At first sight, it appears as if $\gamma$, $r$ and $\delta$ could be determined with
the help of the $U$-spin symmetry from the ratio of the CP-averaged branching
ratios and the two CP asymmetries provided by the $B_d\to\pi^\mp K^\pm$, 
$B_s\to \pi^\pm K^\mp$ system. However, because of the following $U$-spin
relation, which is the counterpart of (\ref{ACP-BR-rel1}), this is actually not the 
case:
\begin{equation}\label{ACP-BR-rel2}
\frac{{\cal A}_{\rm CP}^{\rm dir}(B_s\to\pi^\pm K^\mp)}{{\cal A}_{\rm CP}^{\rm dir}
(B_d\to\pi^\mp K^\pm)}=-\left|\frac{P_s}{P}\right|^2
\left[\frac{M_{B_d}}{M_{B_s}}
\frac{\Phi(M_\pi/M_{B_s},M_K/M_{B_s})}{\Phi(M_\pi/M_{B_d},M_K/M_{B_d})}
\frac{\tau_{B_s}}{\tau_{B_d}}\right]
\left[\frac{\mbox{BR}(B_d\to\pi^\mp K^\pm)}{\mbox{BR}(B_s\to\pi^\pm K^\mp)}\right].
\end{equation}
On the other hand, it allows us to obtain experimental insights into
$U$-spin-breaking effects with the help of the measurements of the 
CP asymmetries and the CP-averaged branching ratios listed in 
Section~\ref{sec:intro}. Adding the errors in quadrature, we obtain
\begin{equation}
\left|\frac{P_s}{P} \right|_{\rm exp}=\left|\frac{P_s}{P} \right|
\sqrt{\Bigl[\frac{r_s}{r}\Bigr]\Bigl[\frac{\sin\delta_s}{\sin\delta}\Bigr]}=1.06\pm0.28,
\end{equation}
where we have also taken non-factorizable $U$-spin-breaking effects to 
(\ref{U-spin-rel-2}) into account. We obtain excellent agreement with 
(\ref{relP-Ps-QCDSR}), although the experimental uncertainties are still large. 
This quantity should be closely monitored as the data improve, allowing us
to obtain valuable insights into non-factorizable $U$-spin-breaking effects. 
We shall return to this issue below.

\boldmath
\subsection{Further Information: $B^+\to\pi^+K^0$ and 
$B^+\to K^+\bar K^0$}\label{ssec:FSI}
\unboldmath
For the determination of $\gamma$ from the $B_d\to\pi^\mp K^\pm$, 
$B_s\to \pi^\pm K^\mp$ system, the overall normalization $P$ has to be fixed
through an additional input, which is offered by the decay $B^+\to \pi^+ K^0$. 
If we neglect colour-suppressed EW penguin topologies and use the 
$SU(2)$ isospin symmetry of strong interactions, we may write its amplitude
as follows:
\begin{equation}\label{ampl-BppipK0}
A(B^+\to \pi^+ K^0)=P\left[1+\epsilon \rho_{\pi K}e^{i\theta_{\pi K}}e^{i\gamma}\right],
\end{equation}
where the CP-conserving hadronic parameter $\rho_{\pi K}e^{i\theta_{\pi K}}$ is 
expected to play a minor r\^ole because of the $\epsilon$ suppression. A first
probe of this quantity is offered by the direct CP asymmetry
\begin{equation}
{\cal A}_{\rm CP}^{\rm dir}(B^\pm\to \pi^\pm K)=
-\left[\frac{2\epsilon\rho_{\pi K}\sin\theta_{\pi K}\sin\gamma}{1
+2\epsilon\rho_{\pi K}\cos\theta_{\pi K}\cos\gamma+\epsilon^2\rho_{\pi K}^2}\right]
=-0.009\pm0.025.
\end{equation}
The experimental value \cite{HFAG}, which is the average of the corresponding 
$B$-factory results, does not indicate any anomalous enhancement of 
$\rho_{\pi K}e^{i\theta_{\pi K}}$. This parameter can actually be determined with 
the help of the $U$-spin-related decay $B^+\to K^+\bar K^0$ \cite{rho-det,RF-BpiK-98}. 
In the SM, its transition amplitude can be 
written as follows:
\begin{equation}
A(B^+\to K^+\bar K^0)=\sqrt{\epsilon}P_{KK} \left[1-\rho_{KK}e^{i\theta_{KK}}
e^{i\gamma}\right],
\end{equation}
where the $U$-spin symmetry implies 
\begin{equation}\label{U-spin-rel-3}
\rho_{KK}= \rho_{\pi K}, \quad \theta_{KK}=\theta_{\pi K}.
\end{equation}
This channel was recently discovered at the $B$ factories with the following
CP-averaged branching ratios:
\begin{equation}\label{BKK-exp}
\mbox{BR}(B^\pm\to K^\pm K)=\left\{
\begin{array}{ll}
(1.61 \pm 0.44 \pm 0.09)\times 10^{-6} & \mbox{(BaBar) \cite{BaBar-BKK}}\\
(1.22^{+0.33+0.13}_{-0.28-0.16})\times10^{-6}&  \mbox{(Belle) \cite{Belle-BKK}},
\end{array}
\right.
\end{equation}
which correspond to the average 
\begin{equation}\label{BKK-av}
\mbox{BR}(B^\pm\to K^\pm K)=\left(1.36^{+0.29}_{-0.27}\right)\times 10^{-6}.
\end{equation}
Moreover, also a first result for the corresponding direct CP asymmetry is 
available:
\begin{equation}\label{ACPdirBKK}
{\cal A}_{\rm CP}^{\rm dir}(B^\pm\to K^\pm K)=
\frac{2\rho_{KK}\sin\theta_{KK}\sin\gamma}{1-2\rho_{KK}\cos\theta_{KK}
\cos\gamma+\rho_{KK}^2}=-0.12_{-0.17}^{+0.18}.
\end{equation}

The branching ratios are interestingly measured close to lower bounds
that can be derived in the SM \cite{FR}. In fact, if we introduce 
\begin{equation}
H_{\pi K}^{KK}\equiv\frac{1}{\epsilon}\left|\frac{P}{P_{KK}}\right|^2
\left[\frac{\Phi(M_\pi/M_B,M_K/M_B)}{\Phi(M_K/M_B,M_K/M_B)}\right]
\left[\frac{\mbox{BR}(B^\pm\to K^\pm K)}{\mbox{BR}(B^\pm\to\pi^\pm K)}\right],
\end{equation}
we obtain
\begin{equation}\label{HKK-expr}
H_{\pi K}^{KK}=\frac{1-2\rho_{KK}\cos\theta_{KK}\cos\gamma+\rho_{KK}^2}{1+
2\epsilon\rho_{\pi K}\cos\theta_{\pi K}\cos\gamma+\epsilon^2\rho_{\pi K}^2}.
\end{equation}
This quantity takes the following lower bound:
\begin{equation}\label{H-bound}
H_{\pi K}^{KK}\geq\left[1-2\epsilon\cos^2\gamma+{\cal O}(\epsilon^2)\right]
\sin^2\gamma,
\end{equation}
which can be converted into a lower bound for $\mbox{BR}(B^\pm\to K^\pm K)$
with the help of the measured $B^\pm\to\pi^\pm K$ branching ratio. Moreover,
also the $U$-spin-breaking corrections to  $|P/P_{KK}|$ have to be determined.
In the factorization approximation, we have
\begin{equation}
\left|\frac{P_{KK}}{P}\right|_{\rm fact}=\frac{F_{BK}(M_K^2;0^+)}{F_{B\pi}(M_K^2;0^+)}
\left(\frac{M_B^2-M_K^2}{M_B^2-M_\pi^2}\right).
\end{equation}
Using once again the QCD sum-rule results of Ref.~\cite{KMM} yields
\begin{equation}
\left|\frac{P_{KK}}{P}\right|_{\rm fact}^{\rm QCDSR}=1.35^{+0.11}_{-0.09},
\end{equation}
which agrees with an alternative analysis \cite{BZ}. The experimental branching 
ratio in (\ref{BR-BppipK}) and the result for $\gamma$ in (\ref{BsKKBdpipi-2}) 
yield then the following lower bound:
\begin{equation}
\mbox{BR}(B^\pm\to K^\pm K)_{\rm min}=\left(1.78^{+0.23}_{-0.26}\right)
\times10^{-6},
\end{equation}
where all errors were again added in quadrature. While the BaBar result in
(\ref{BKK-exp}) is fully consistent with this bound, the Belle measurement
is clearly on the lower side, and reduces also the average in (\ref{BKK-av}),
which yields
\begin{equation}\label{HKK-res}
H_{\pi K}^{KK}=0.64\pm0.15.
\end{equation}
Using (\ref{H-bound}), this value can be converted into the following upper
bound on $\gamma$:
\begin{equation}
\gamma\leq\left(53^{+10}_{-9}\right)^\circ.
\end{equation}
It is about $1\,\sigma$ below the result for $\gamma$ in (\ref{BsKKBdpipi-2}),
which is another manifestation of the low branching ratio in (\ref{BKK-av}).

\begin{figure}
\centerline{
 \includegraphics[width=7.5truecm]{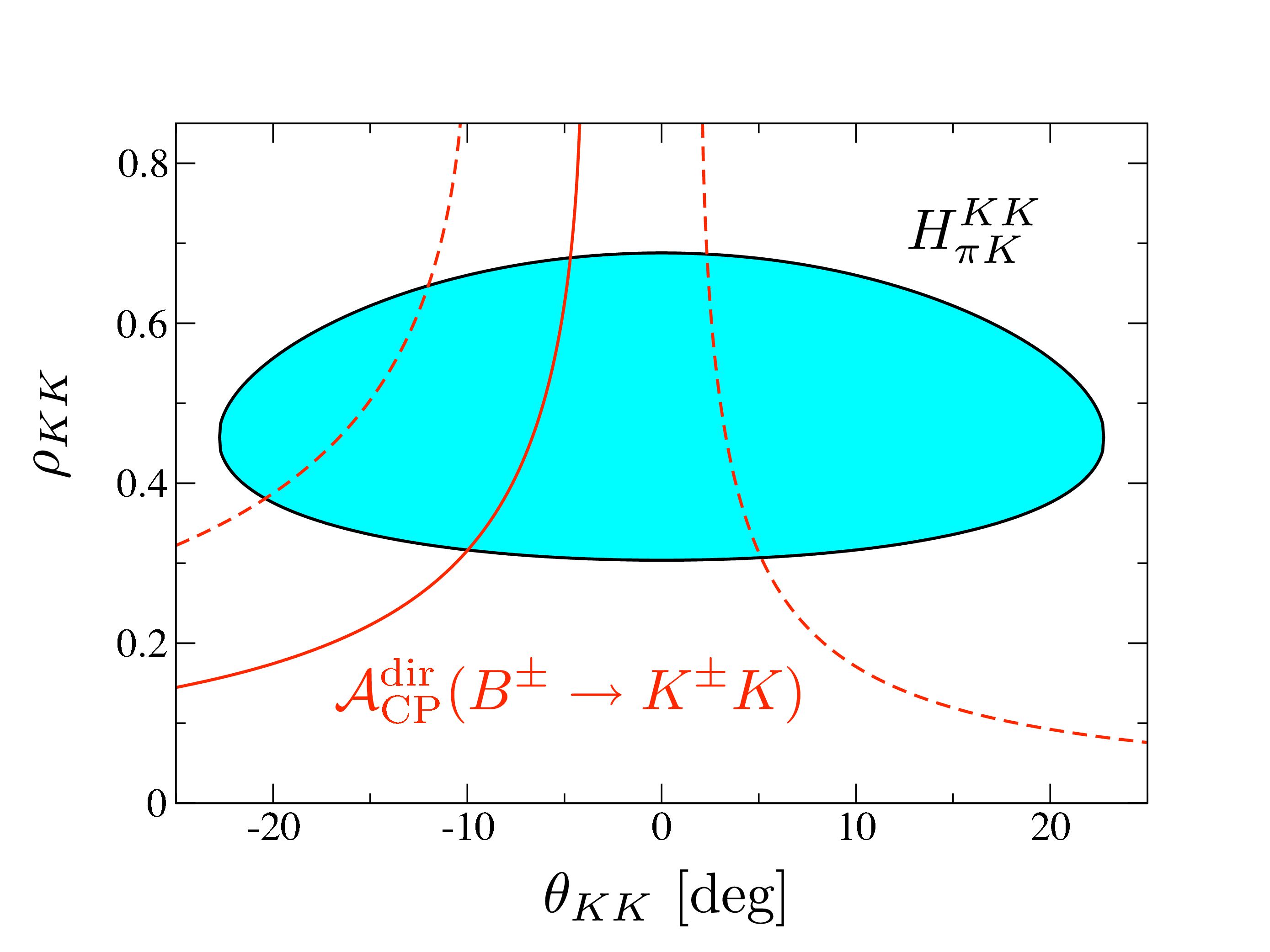}}
 \vspace*{-0.3truecm}
\caption{The constraints in the $\theta_{KK}$--$\rho_{KK}$ plane plane 
following from $H_{\pi K}^{KK}$ and ${\cal A}_{\rm CP}^{\rm dir}(B^\pm\to K^\pm K)$,
as explained in the text ($\gamma=61.6^\circ$, 
$H_{\pi K}^{KK}=0.79$).}\label{fig:rho-KK}
\end{figure}

For a given value of $\gamma$,  (\ref{HKK-expr}) allows us to 
calculate $\rho_{KK}$ as a function of $\theta_{KK}$ with the help
of the $U$-spin relations in (\ref{U-spin-rel-3}):
\begin{equation}\label{rho-thet}
\rho_{KK}=a\pm\sqrt{a^2-b},
\end{equation}
where
\begin{equation}\label{a-b}
a=\left[\frac{1+\epsilon H_{\pi K}^{KK}}{1-\epsilon^2H_{\pi K}^{KK}}\right]
\cos\theta_{KK}\cos\gamma, \quad
b=\frac{1-H_{\pi K}^{KK}}{1-\epsilon^2H_{\pi K}^{KK}}.
\end{equation}
Another contour can be fixed through the direct CP asymmetry in  
(\ref{ACPdirBKK}). To this end, we have just to make the following
replacements in (\ref{rho-thet}):
\begin{equation}
a\to \cos\gamma\cos\theta_{KK}+
\frac{\sin\gamma\sin\theta_{KK}}{{\cal A}_{\rm CP}^{\rm dir}(B^\pm\to K^\pm K)}, 
\quad b\to 1.
\end{equation}
It should be emphasized that this curve is valid exactly, i.e.\ does not
rely on the $U$-spin symmetry. Because of the bounds discussed above, 
(\ref{rho-thet}) with (\ref{a-b}) does not give physical solutions for the 
central values of $\gamma=66.6^\circ$ and $H_{\pi K}^{KK}=0.64$. However, 
if we lower $\gamma$ by one sigma to $61.6^\circ$ and increase
$H_{\pi K}^{KK}$ by one sigma to $0.79$, we arrive at the situation shown in 
Fig.~\ref{fig:rho-KK}, leaving us with a pretty constrained allowed region around
\begin{equation}\label{rhoKK-constr}
\rho_{KK}\approx \rho_{\pi K}\sim0.5, \quad
\theta_{KK}\approx\theta_{\pi K}\sim 0^\circ.
\end{equation}
Consequently, we find $\epsilon\rho_{\pi K}|_{\rm exp}\sim 0.025$, so that
we do not have to worry about the effects of this parameter. In toy models of 
final-state interaction effects that were considered several years ago,
this parameter would have been enhanced by up to one order of magnitude. 
These scenarios are therefore ruled out by the $B$-factory data. Moreover, 
anomalous enhancements of colour-suppressed EW penguin contributions, 
which would arise in such scenarios as well, are also disfavoured.

\boldmath
\subsection{Extracting the UT Angle $\gamma$}\label{ssec:gam-r-del}
\unboldmath
Let us first have a look at the $B_d\to\pi^\mp K^\pm$, $B^\pm\to \pi^\pm K$ system.
For the extraction of $\gamma$, we introduce the following ratio \cite{FM}:
\begin{equation}\label{R-def}
R\equiv\left[\frac{M_{B_d}}{M_{B^+}}
\frac{\Phi(M_\pi/M_{B^+},M_K/M_{B^+})}{\Phi(M_\pi/M_{B_d},M_K/M_{B_d})}
\frac{\tau_{B^+}}{\tau_{B_d}}\right]
\left[\frac{\mbox{BR}(B_d\to\pi^\mp K^\pm)}{\mbox{BR}(B^\pm\to\pi^\pm K)}
\right]=0.899\pm0.049,
\end{equation}
where we have included tiny phase-space effects, used 
$\tau_{B^+}/\tau_{B^0_d}=1.071\pm0.009$ \cite{PDG}, and
added the errors in quadrature. The amplitude parametrizations in 
(\ref{B0pimKp-ampl}) and (\ref{ampl-BppipK0}) imply then the following 
expression \cite{RF-BpiK-98}:
\begin{equation}\label{R-expr}
w^2 R =1-2r\cos\delta\cos\gamma+r^2,
\end{equation}
with
\begin{equation}
w=\sqrt{1+2\epsilon\rho_{\pi K}\cos\theta_{\pi K}+\epsilon^2\rho_{\pi K}^2}.
\end{equation}
Using (\ref{rhoKK-constr}), we obtain $w^2\sim 1.02$. The corresponding effect
lies within the errors of (\ref{R-def}) and will be neglected in the following discussion. 
Following Ref.~\cite{FM}, where the bound
\begin{equation}\label{FM-bound}
\sin^2\gamma \leq R
\end{equation}
was derived, we obtain
\begin{equation}\label{FM-bound-num}
\gamma\leq \left(71.5^{+5.3}_{-4.3}\right)^\circ,
\end{equation}
where the errors reflect the uncertainties of $R$. The value of $\gamma$
in (\ref{BsKKBdpipi-2}) and the SM fits of the UT are well consistent with
this bound, which effectively constrains $\gamma$ in a phenomenologically
very interesting region.

If we combine $R$ with the direct CP asymmetry of $B^0_d\to\pi^-K^+$, 
the strong phase $\delta$ can be eliminated, allowing us to calculate
$r$ as a function of $\gamma$. To this end, it is convenient to introduce
the following ``pseudo-asymmetry" \cite{GR-98}:
\begin{equation}\label{A0-def}
A_0\equiv{\cal A}_{\rm CP}^{\rm dir}(B_d\to\pi^\mp K^\pm) R=2r\sin\delta\sin\gamma,
\end{equation}
so that
\begin{equation}\label{r-det}
r=\sqrt{a_d\pm\sqrt{a_d^2-b_d}},
\end{equation}
with
\begin{eqnarray}
a_d&=&R-\sin^2\gamma+\cos^2\gamma,\\
b_d&=&(1-R)^2+\left(\frac{A_0\cos\gamma}{\sin\gamma}\right)^2;
\end{eqnarray}
for generalized expressions, taking also the effects of $(\rho_{\pi K},\theta_{\pi K})$
and colour-suppressed EW penguins into account, see Ref.~\cite{RF-BpiK-98}.
For given values of $\gamma$ and $r$, the strong phase $\delta$ can unambiguously
be determined through
\begin{eqnarray}
r\cos\delta&=&\cos\gamma\pm\mbox{sgn}(\cos\gamma)\sqrt{\cos^2\gamma-
(1-R)-\left(\frac{A_0}{2\sin\gamma}\right)^2},\\
r\sin\delta&=&\frac{A_0}{2 \sin\gamma}.
\end{eqnarray}
As $R<1$, we have $\mbox{sgn}(\cos\delta)=\mbox{sgn}(\cos\gamma)$ for
the two solutions of $r$. Consequently, since we expect a positive value of
the cosine of $\delta$, as in factorization, we are left with the range of 
$-90^\circ<\gamma<+90^\circ$. Since the four solutions for $\gamma$
following from (\ref{BsKKBdpipi-1}) and (\ref{BsKKBdpipi-2}) with 
(\ref{ambig-1}) overlap with that region only for $0^\circ<\gamma<90^\circ$, 
we may restrict the following discussion to this range.

\begin{figure}
\centerline{
\begin{tabular}{cc}
 \includegraphics[width=7.5truecm]{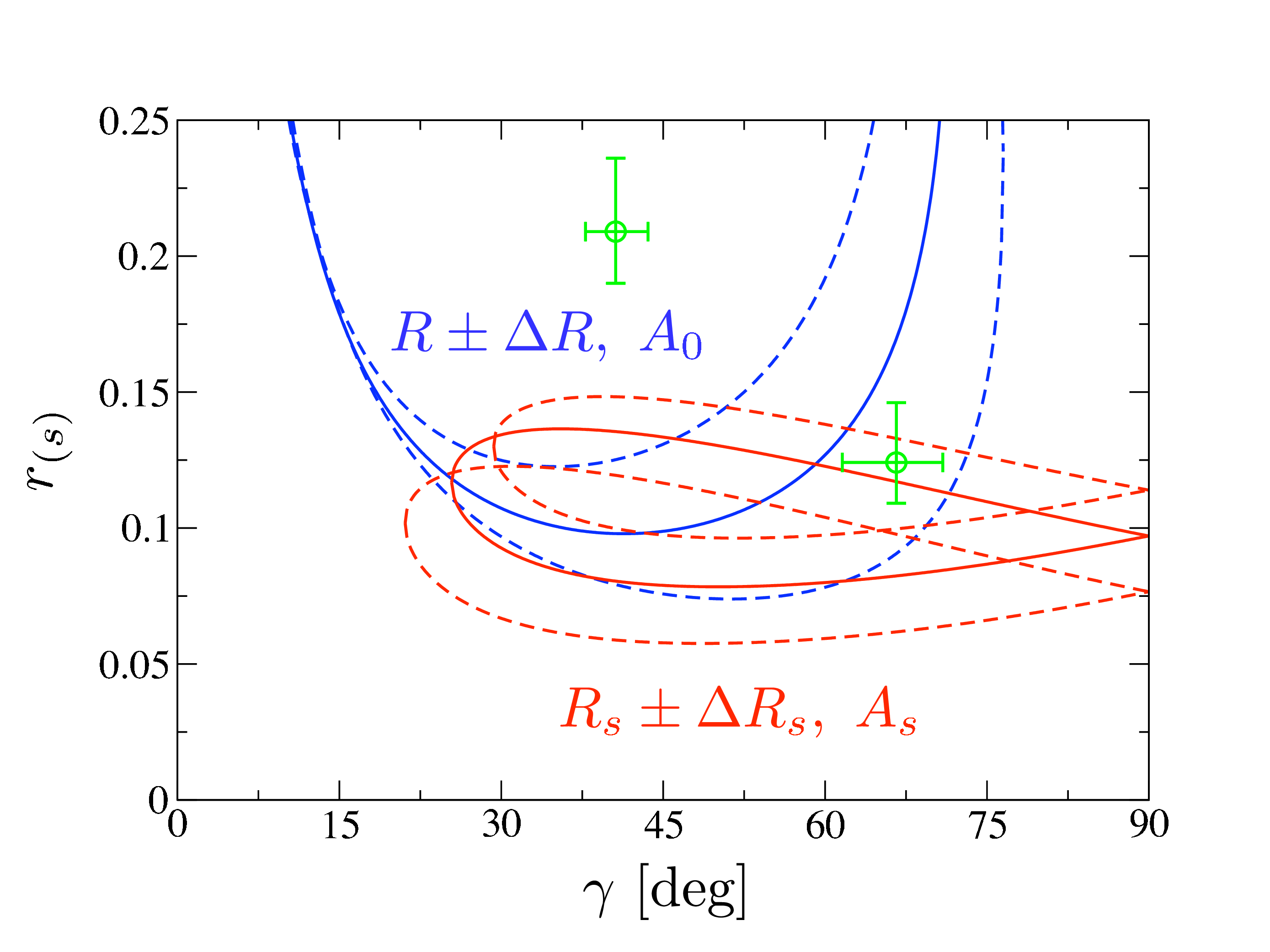} &
 \includegraphics[width=7.5truecm]{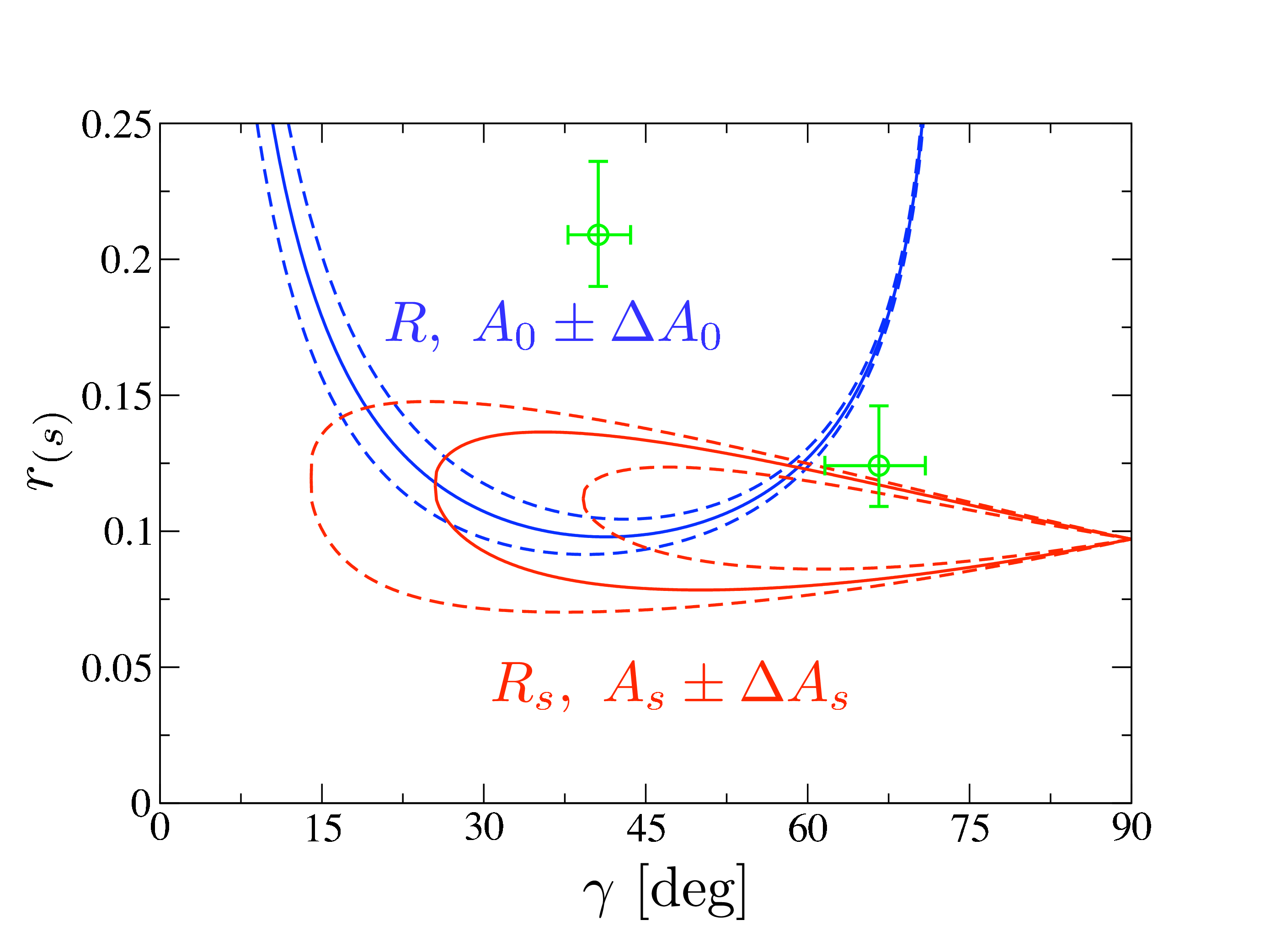}
 \end{tabular}}
 \vspace*{-0.3truecm}
\caption{The contours in the $\gamma$--$r_{(s)}$ plane: the left panel
shows the $1\,\sigma$ ranges of the $R_{(s)}$ (upper and lower curves
correspond to $R_{(s)}+\Delta R_{(s)}$ and  $R_{(s)}-\Delta R_{(s)}$, 
respectively), the right panel the $1\,\sigma$ ranges of the corresponding 
direct CP asymmetries (upper and lower curves
correspond to ${\cal A}_{\rm CP}^{\rm dir}+\Delta {\cal A}_{\rm CP}^{\rm dir}$
and ${\cal A}_{\rm CP}^{\rm dir}-\Delta {\cal A}_{\rm CP}^{\rm dir}$, respectively).
The error bars represent the results for $\gamma$ and $d$ in 
(\ref{BsKKBdpipi-1}) and (\ref{BsKKBdpipi-2}).}\label{fig:R-cont-1}
\end{figure}

The determination of $\gamma$ requires further information, which can be
obtained with the help of the $B^0_s\to\pi^+K^-$ channel. To this end, 
we introduce -- in analogy to (\ref{R-def}) and (\ref{A0-def}) -- the following 
quantities:
\begin{eqnarray}
R_s&\equiv&\left|\frac{P}{P_s}\right|^2
\left[\frac{M_{B_s}}{M_{B^+}}
\frac{\Phi(M_\pi/M_{B^+},M_K/M_{B^+})}{\Phi(M_\pi/M_{B_s},M_K/M_{B_s})}
\frac{\tau_{B^+}}{\tau_{B_s}}\right]
\left[\frac{\mbox{BR}(B_s\to\pi^\pm K^\mp)}{\mbox{BR}(B^\pm\to\pi^\pm K)}
\right]\nonumber\\
&=&\epsilon + 2 r_s\cos\delta_s\cos\gamma+\frac{r_s^2}{\epsilon}
=0.236\pm0.070,\label{Rs-def}
\end{eqnarray}
where (\ref{relP-Ps-QCDSR}) as well as $\tau_{B^+}=(1.638\pm0.011)\mbox{ps}$
and $\tau_{B_s}=(1.466\pm0.059)\mbox{ps}$ \cite{PDG} enter the numerical 
value, and
\begin{equation}\label{A0s-def}
A_s\equiv{\cal A}_{\rm CP}^{\rm dir}(B_s\to\pi^\pm K^\mp) R_s=-2r_s\sin\delta_s
\sin\gamma.
\end{equation}
These quantities allow us to eliminate the strong phase $\delta_s$, and to 
calculate $r_s$ as a function of $\gamma$. To this end, we have simply to 
make the replacements $r\to r_s$, $a_d\to a_s$ and $b_d\to b_s$ in (\ref{r-det}), 
with
\begin{eqnarray}
a_s&=&\epsilon\left[R_s-\epsilon\left(\sin^2\gamma-\cos^2\gamma\right)\right],\\
b_s&=&\epsilon^2\left[(R_s-\epsilon)^2+\left(\frac{A_s\cos\gamma}{\sin\gamma}
\right)^2\right].
\end{eqnarray}
For given values of $\gamma$ and $r_s$, we may again extract the strong phase
unambiguously with the help of the relations
\begin{eqnarray}
r_s\cos\delta_s&=&-\epsilon\cos\gamma\mp\mbox{sgn}(\cos\gamma)
\sqrt{\epsilon\left(R_s-\epsilon\sin^2\gamma\right)-\left(\frac{A_s}{2\sin\gamma}
\right)^2},\label{del-s}\\
r_s\sin\delta_s&=&-\left(\frac{A_s}{2 \sin\gamma}\right).
\end{eqnarray}

Using, finally, the first $U$-spin relation given in (\ref{U-spin-rel-2}), the 
intersection of the $\gamma$--$r$ and $\gamma$--$r_s$ contours allows 
the extraction of $\gamma$ and $r_s=r$. Moreover, also the strong
phases can be extracted, providing an internal consistency check of the
$U$-spin symmetry; $U$-spin-breaking corrections to $r_s=r$ correspond
to a relative shift of both contours. In Fig.~\ref{fig:R-cont-1}, we show 
these curves for the current data, exploring also the impact of their uncertainties. 
The realization of the bound in (\ref{FM-bound}) is nicely visible. On the other 
hand, the contour plots show also that the situation for the extraction of
$\gamma$ is not as fortunate as in the case of the 
$B_d\to\pi^+\pi^-$, $B_s\to K^+K^-$ system discussed in Section~\ref{sec:BsKK}.
Moreover, further constraints arise from $\cos\delta_s$, which has to 
agree with the positive sign of $\cos\delta$. A closer look at (\ref{del-s}) shows,
that this is only the case for the lower branches of the $\gamma$--$r_s$ contours,
i.e.\ for the minus (plus) signs in (\ref{r-det}) ((\ref{del-s})). Combining all this
information, we arrive at the following ranges:
\begin{equation}\label{BspiK-range}
26^\circ\leq\gamma\leq70^\circ,  \quad 0.07 \leq r\leq0.12.
\end{equation}

Since only the lower branches of the $\gamma$--$r_s$ contours are effective
because of the constraint on $\delta_s$, a solution for $\gamma$ around $65^\circ$
requires in particular an increase of $R_s$, which still suffers from significant 
uncertainties, and would welcome an increase of $R$ as well, which is known
at the $5\%$ level. In principle, such an effect could be due to the $\rho_{\pi K}$ 
parameter in (\ref{ampl-BppipK0}). However, the analysis of 
Subsection~\ref{ssec:FSI} demonstrates that this corresponds to only a few percent. 
Interestingly, it shifts $R$ in the right direction, but this effect is definitely much 
too small to cure the problem with $R_s$. If we consider the upper $1\,\sigma$ 
values of $R_s=0.306$ and $R=0.948$, we obtain the following values:
\begin{equation}
\gamma=69.4^\circ, \quad r=0.101, \quad \delta=28.5^\circ, \quad
\delta_s=39.2^\circ,
\end{equation}
which would look quite reasonable. 

Using both $U$-spin relations in (\ref{U-spin-rel-2}) simultaneously, 
the following expression can straightforwardly be derived from (\ref{R-expr}) 
and (\ref{Rs-def}):
\begin{equation}
r=\sqrt{\epsilon\left[\frac{R+R_s-1-\epsilon}{1+\epsilon}\right]}.
\end{equation}
In the case of the central values of the current experimental results, 
we obtain $r=0.06$, whereas  the upper $1\,\sigma$ values yield $r=0.10$. 
The advantage of the contours in the $\gamma$--$r_{(s)}$ plane is that the 
strong phases $\delta$ and $\delta_s$ can be extracted separately.

\boldmath
\subsection{Interplay with the $B_s\to K^+K^-$, $B_d\to\pi^+\pi^-$
Strategy}\label{ssec:contact}
\unboldmath
If we replace the strange spectator quark of $B^0_s\to K^+K^-$ through a 
down quark, we obtain the $B^0_d\to\pi^-K^+$ decay, as can be seen
in Fig.~\ref{fig:U-spin-diag}. Consequently, the only difference between
the corresponding hadronic matrix elements is due to processes involving
these spectator quarks: penguin annihilation and exchange topologies,
which contribute to $B^0_s\to K^+K^-$, but are absent in the
$B^0_d\to\pi^-K^+$ channel. These contributions, which are expected to
play a minor r\^ole,  can be probed through $B_d\to K^+K^-$ and 
$B_s\to \pi^+\pi^-$ decays \cite{GHLR}. The most recent data for the 
corresponding CP-averaged branching ratios read as follows \cite{HFAG}:
\begin{eqnarray}
\mbox{BR}(B_d\to K^+ K^-) & =& \left(0.15^{+0.11}_{-0.10}\right)\times10^{-6},  \\
\mbox{BR}(B_s\to \pi^+\pi^-) & = & \left(0.53\pm0.51\right)\times10^{-6},
\end{eqnarray}
where the constraint on the $B_s$ mode was recently obtained
at the Tevatron \cite{CDF-punzi}.
Following Ref.~\cite{BFRS}, these measurements can be converted into 
constraints on strong amplitudes:
\begin{eqnarray}
\lefteqn{\sqrt{\frac{1}{2}\left[\frac{\mbox{BR}(B_d\to K^+K^-)}{\mbox{BR}(B^\pm
\to\pi^\pm\pi^0)}\right]\frac{\tau_{B^+}}{\tau_{B_d}}}}\nonumber\\
&&\approx
\left|\frac{{\cal E}-({\cal PA})_{tu}}{{\cal T+C}}\right|
\sqrt{1+2\varrho_{{\cal PA}}\cos\vartheta_{{\cal PA}}\cos\gamma+\varrho_{{\cal PA}}^2}
=0.12^{+0.04}_{-0.06},
\end{eqnarray}
\begin{equation}
\sqrt{\frac{\epsilon}{2}\left[\frac{\mbox{BR}(B_s\to \pi^+\pi^-)}{\mbox{BR}(B^\pm
\to\pi^\pm\pi^0)}\right]\frac{\tau_{B^+}}{\tau_{B_s}}}\approx
\frac{1}{R_b}\left|\frac{({\cal PA})_{tc}}{{\cal T+C}}\right|= 0.05^{+0.03}_{-0.04}.
\end{equation}
Here ${\cal T+C}$ describes the sum of colour-allowed and colour-suppressed tree 
topologies, ${\cal E}$ is an exchange amplitude, whereas the $({\cal PA})_{tq}$ 
are the differences of penguin annihilation amplitudes with internal top and 
$q\in\{u,c\}$ quarks. Finally, 
\begin{equation}
\varrho_{{\cal PA}}e^{i\vartheta_{{\cal PA}}}\equiv\frac{1}{R_b}\left[
\frac{({\cal PA})_{tc}}{{\cal E}-({\cal PA})_{tu}}\right],
\end{equation}
with $R_b\approx0.4$ denoting the side of the UT that is proportional to 
$|V_{ub}/V_{cb}|$. Consequently, the data from the $B$ factories and the
Tevatron do not indicate any anomalous behaviour of these topologies so that
we will neglect them in the following discussion. Similar assumptions were made
in the recent extractions of $\gamma$ from $B_d\to\pi^+\pi^-$, $B\to\pi K$ modes
in Refs.~\cite{FRS,GR-07}, yielding results that agree within the errors with 
our value of $\gamma$ in (\ref{BsKKBdpipi-2-U}).

\begin{figure}
\centerline{
 \includegraphics[width=7.5truecm]{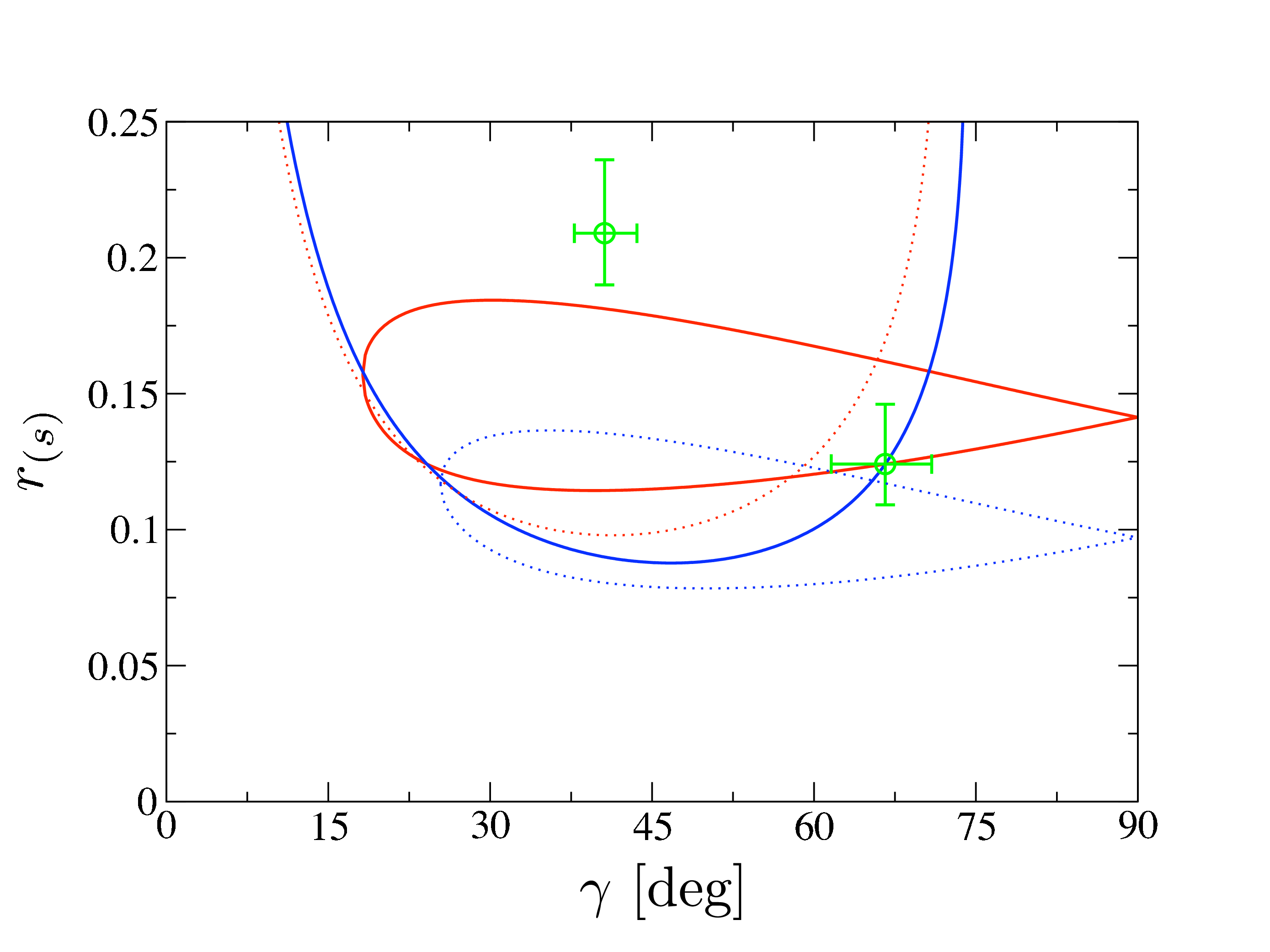}}
 \vspace*{-0.3truecm}
\caption{Future scenario for the contours in the $\gamma$--$r_{(s)}$ plane,
as discussed in the text. The dotted lines refer to the central values of the
current data.}\label{fig:R-cont-scen}
\end{figure}

Applying the $SU(3)$ flavour symmetry, we may then identify the $B^0_s\to K^+K^-$ 
and $B^0_d\to\pi^-K^+$ decay amplitudes \cite{RF-BsKK,RF-00,FlMa-2}, and
obtain the simple relation:
\begin{equation}\label{r-d-rel}
r e^{i\delta}=\frac{\epsilon}{d}e^{i(\pi-\theta)},
\end{equation}
which allows us to convert (\ref{BsKKBdpipi-1}) and (\ref{BsKKBdpipi-2}) 
into their $B^0_d\to\pi^-K^+$ counterparts:
\begin{equation}\label{BsKKBdpipi-1-conv}
\gamma=(40.6^{+3.0}_{-2.8})^\circ, \quad
r=0.209^{+0.027}_{-0.019}, \quad
\delta=(150.8_{-15.3}^{+13.3})^\circ,
\end{equation}
\begin{equation}\label{BsKKBdpipi-2-conv}
\gamma = (66.6^{+4.3}_{-5.0})^\circ, \quad
r = 0.124^{+0.022}_{-0.015}, \quad
\delta=(24.1_{-11.1}^{+4.5})^\circ.
\end{equation}
In Fig.~\ref{fig:R-cont-1}, we have included these values as the two points
with error bars. We can nicely see that the $\gamma$--$r$ contour, which 
is fixed through $R$ and $A_0$, clearly rules out (\ref{BsKKBdpipi-1-conv}),
as we noted in Subsection~\ref{ssec:ambig}. So we are left with the SM-like 
solution of (\ref{BsKKBdpipi-2-conv}), which would favour a slight increase
of $R$, and quite a significant increase of $R_s$. In fact, if we calculate
these quantities for that case, we obtain
\begin{equation}
R=0.925^{+0.018}_{-0.021}, 
\quad R_s=0.444^{+0.137}_{-0.084},
\end{equation}
where the errors are due to our input parameters. Converting the value of
$R_s$ into the $B_s\to\pi^\pm K^\mp$ branching ratio yields
\begin{equation}\label{Bs-pred-1}
\mbox{BR}(B_s\to\pi^\pm K^\mp)=\left(9.4^{+3.3}_{-2.3}\right)\times 10^{-6},
\end{equation}
which is about $1.6\,\sigma$ larger than the CDF result in (\ref{BR-BspiK}). 
The prediction of the direct CP violation in $B_d\to\pi^\mp K^\pm$ yields
\begin{equation}\label{CPdir-pred}
{\cal A}_{\rm CP}^{\rm dir}(B_d\to\pi^\mp K^\pm)=+0.101^{+0.055}_{-0.047},
\end{equation}
with the same numerical value as the prediction of 
${\cal A}_{\rm CP}^{\rm dir}(B_s\to K^+K^-)$ in Subsection~\ref{ssec:ACPmix}.
In fact, using the assumptions listed above, we expect
\begin{equation}\label{ACP-BsKK-exp}
{\cal A}_{\rm CP}^{\rm dir}(B_s\to K^+K^-)=
{\cal A}_{\rm CP}^{\rm dir}(B_d\to\pi^\mp K^\pm)\stackrel{\rm exp}{=}0.095\pm0.013.
\end{equation}
Moreover, we have ${\cal A}_{\rm CP}^{\rm dir}(B_s\to\pi^\pm K^\mp)=-0.21$, which
is equal to our input parameter for the direct CP asymmetry of the
$B_d\to\pi^+\pi^-$ channel. The agreement between (\ref{CPdir-pred}) and the
experimental value in (\ref{ACP-BsKK-exp}) is remarkable, and disfavours
large $SU(3)$-breaking corrections, in particular to the relations between
strong phases (see Subsection~\ref{ssec:ACPmix}). In Fig.~\ref{fig:R-cont-scen},
we show the corresponding situation in the $\gamma$--$r_{(s)}$ plane as a
future scenario for the evolution of the data. 

Let us finally return to the CP-averaged branching ratios, where the relation
\begin{equation}\label{BR-rel-11}
\frac{\mbox{BR}(B_s\to K^+K^-)}{\mbox{BR}(B_d\to\pi^\mp K^\pm)}=
\left[\frac{M_{B_d}}{M_{B_s}}
\frac{\Phi(M_K/M_{B_s},M_K/M_{B_s})}{\Phi(M_\pi/M_{B_d},M_K/M_{B_d})}
\frac{\tau_{B_s}}{\tau_{B_d}}\right]\left(\frac{f_\pi}{f_K}\left|\frac{{\cal C}'}{{\cal C}}
\right|_{\rm fact}\right)^2
\end{equation}
allows us to extract
\begin{equation}\label{CpC-exp}
\left|\frac{{\cal C}'}{{\cal C}}\right|_{\rm fact}^{\rm exp}=1.42\pm0.14
\end{equation}
from the data. Within the uncertainties, this number agrees remarkably
well with (\ref{SR-1}), and gives us further confidence into the corresponding
form factors and the smallness of non-factorizable $SU(3)$-breaking effects.
In analogy to (\ref{BR-rel-11}), we also have
\begin{equation}\label{BR-rel-22}
\frac{\mbox{BR}(B_s\to \pi^\pm K^\pm)}{\mbox{BR}(B_d\to\pi^+\pi^-)}=
\left[\frac{M_{B_d}}{M_{B_s}}\frac{\Phi(M_\pi/M_{B_s},M_K/M_{B_s})}{\Phi(M_\pi/M_{B_d},M_\pi/M_{B_d})}\frac{\tau_{B_s}}{\tau_{B_d}}\right]
\left(\frac{f_K}{f_\pi}\left|\frac{P_s}{P}\right|_{\rm fact}\right)^2.
\end{equation}
Using the numerical value in (\ref{relP-Ps-QCDSR}) with $f_K/f_\pi=1.22$, we obtain
\begin{equation}\label{BspiK-pred-FF}
\mbox{BR}(B_s\to \pi^\pm K^\pm)=\left(7.5\pm1.2\right)\times10^{-6}.
\end{equation}
This prediction is a bit smaller than (\ref{Bs-pred-1}), but fully consistent within
the errors. On the other hand, it is about $1.4\,\sigma$ larger than the experimental
value in (\ref{BR-BspiK}), thereby giving further support for the observations 
made above and in Subsection~\ref{ssec:gam-r-del}. Using the $U$-spin relation
in (\ref{ACP-BR-rel2}), the enhancement of the central value of 
$\mbox{BR}(B_s\to \pi^\pm K^\pm)$ by a factor of $1.5$ would suppress
the central value of (\ref{ACP-Bs-exp}) to
\begin{equation}
{\cal A}_{\rm CP}^{\rm dir}(B_s\to\pi^\pm K^\mp)\sim-0.26,
\end{equation}
which would further support the BaBar measurement in (\ref{ACP-dir-pipi-ex}), as
\begin{equation}
{\cal A}_{\rm CP}^{\rm dir}(B_s\to\pi^\pm K^\mp)\approx
{\cal A}_{\rm CP}^{\rm dir}(B_d\to\pi^+\pi^-).
\end{equation}
Since the form-factor ratio
\begin{equation}
\frac{f_\pi}{f_K}\left|\frac{{\cal C}'}{{\cal C}}
\right|_{\rm fact}=\frac{F_{B_sK}(M_K^2;0^+)}{F_{B_d\pi}(M_\pi^2;0^+)}
\left(\frac{M_{B_s}^2-M_K^2}{M_{B_d}^2-M_\pi^2}\right)
\end{equation}
is essentially equal to
\begin{equation}
\frac{f_K}{f_\pi}\left|\frac{P_s}{P}\right|_{\rm fact}=
\frac{F_{B_sK}(M_\pi^2;0^+)}{F_{B_d\pi}(M_K^2;0^+)}
\left(\frac{M_{B_s}^2-M_K^2}{M_{B_d}^2-M_\pi^2}\right),
\end{equation}
we arrive at the following relation, which does not depend on the
form-factor ratios:\footnote{In (\ref{BR-rel-11}) and (\ref{BR-rel-22}), actually 
$F_{B_d\pi}(M_K^2;0^+)$ and $F_{B_d\pi}(M_\pi^2;0^+)$ enter, respectively. }
\begin{equation}
\mbox{BR}(B_s\to \pi^\pm K^\pm)=
\left[\frac{\mbox{BR}(B_s\to K^+K^-)}{\mbox{BR}(B_d\to\pi^\mp K^\pm)}\right]
\mbox{BR}(B_d\to\pi^+\pi^-)=(6.5\pm1.3)\times10^{-6}.
\end{equation}
If we increase $\mbox{BR}(B_s\to K^+K^-)$ by a factor of $1.15$ in order to 
get full agreement between the central values of (\ref{CpC-exp}) and (\ref{SR-1})
(which is below a $1\,\sigma$ fluctuation and would have a small impact
on the $\gamma$ determination in Section~\ref{sec:BsKK}), we would 
arrive again at (\ref{BspiK-pred-FF}). 
Instead of predicting this branching ratio, we may perform an experimental test
of non-factorizable $SU(3)$-breaking effects:
\begin{equation}
\Delta^{\rm NF}_{SU(3)}\equiv
1-\left[\frac{\mbox{BR}(B_s\to K^+K^-)}{\mbox{BR}(B_s\to \pi^\pm K^\pm)}
\right]\left[\frac{\mbox{BR}(B_d\to\pi^+\pi^-)}{\mbox{BR}(B_d\to\pi^\mp K^\pm)}\right]
=-0.3\pm0.4.
\end{equation}
In view of the large uncertainties, this relation is not yet very constraining. However,
it should provide valuable insights as the data improve.

\section{Conclusions}\label{sec:concl}
\setcounter{equation}{0}
We have performed an analysis of the $U$-spin-related decays
$B_d\to\pi^+\pi^-$, $B_s\to K^+K^-$ and $B_d\to\pi^\mp K^\pm$,
$B_s\to\pi^\pm K^\mp$, exploring the implications of the current
$B$-factory data and the first results on the $B_s$ modes from the
Tevatron and setting the stage for the data taking at the LHC. The 
main results can be summarized as follows:
\begin{itemize}
\item The analysis of the $B_d\to\pi^+\pi^-$, $B_s\to K^+K^-$ system
favours the BaBar measurement of the direct CP violation in the 
former decay. We have performed the first determination of $\gamma$
by using only $U$-spin-related decays, and found a particularly 
fortunate situation, yielding $\gamma=(66.6^{+4.3+4.0+0.1}_{-5.0-3.0-0.2})^\circ$,
where the first errors reflect the uncertainties of the input quantities, and
the second and third errors show the sensitivity to generous 
non-factorizable $U$-spin-breaking corrections.
\item This value of $\gamma$ is in excellent agreement with the SM
fits of the UT. We have shown how discrete ambiguities can be resolved 
through ${\cal A}_{\rm CP}^{\rm mix}(B_s\to K^+K^-)$, which has not yet 
been measured. However, may use alternatively the observables of the  
$B_d\to\pi^\mp K^\pm$, $B^\pm\to\pi^\pm K$ modes, leaving us with the 
result for $\gamma$ given above.
\item The next important step in this analysis will be the observation of 
mixing-induced CP violation in the $B^0_s\to K^+K^-$ decay. In the
SM, we predict this asymmetry as ${\cal A}_{\rm CP}^{\rm mix}(B_s\to K^+K^-)=
-0.246^{+0.036+0.008+0.051}_{-0.030-0.007-0.023}$, where the second and
third errors illustrate again the impact of large non-factorizable 
$U$-spin-breaking corrections. We have also explored the impact of CP-violating
NP contributions to $B^0_s$--$\bar B^0_s$ mixing on this observable, which
affect it sensitively. Moreover, we pointed out that the measurements
of ${\cal A}_{\rm CP}^{\rm mix}(B_s\to K^+K^-)$  and $\sin\phi_s$ through
$B_s\to J/\psi\phi$ will allow an {\it unambiguous} determination of 
the $B^0_s$--$\bar B^0_s$ mixing phase $\phi_s$.
\item Using the results of our analysis, the measurement of the
$B_s\to K^+K^-$ lifetime through an untagged data sample can be converted
into the width difference $\Delta\Gamma_s$. In the SM, the corresponding
key observable is given by
${\cal A}_{\Delta\Gamma}(B_s\to K^+K^-)=-0.964^{+0.011}_{-0.007}$,
which is essentially unaffected by $U$-spin-breaking corrections. 
\item In the case of the $B_d\to\pi^\mp K^\pm$, $B_s\to\pi^\pm K^\mp$
system, the determination of $\gamma$ requires an additional input, 
which is provided by $B^\pm\to\pi^\pm K$. In contrast to the $B_d\to\pi^+\pi^-$, 
$B_s\to K^+K^-$ system, we have then also to make additional dynamical
assumptions. In particular, another hadronic parameter enters $B^\pm\to\pi^\pm K$,
which is doubly Cabibbo-suppressed, but could be enhanced by final-state 
interaction effects. Using the $B$-factory data for $B^\pm\to K^\pm K$ modes, 
we have shown that this is actually not the case, and that these effects can 
safely be neglected. This does also support the neglect of colour-suppressed
EW penguins.
\item Using $\mbox{BR}(B^\pm\to\pi^\pm K)$ to normalize the 
branching ratios of $B_d\to\pi^\mp K^\pm$ and $B_s\to \pi^\pm K^\mp$, we 
have introduced two quantities $R$ and $R_s$, respectively. In the case of
$R$, the bound of $\gamma\leq\left(71.5^{+5.3}_{-4.3}\right)^\circ$ is implied,
which puts a constraint on this UT angle in a phenomenologically interesting
region. If we combine $R$ and $R_s$ with the direct CP asymmetries of
the $B_d\to\pi^\mp K^\pm$ and $B_s\to \pi^\pm K^\mp$ modes, respectively,
we can extract $\gamma$, a hadronic parameter $r$, and two strong phases
with the help of the $U$-spin symmetry. The situation resulting from the
current data leaves us with $26^\circ\leq\gamma\leq70^\circ$, and is not
as favourable as in the case of $B_d\to\pi^+\pi^-$, $B_s\to K^+K^-$.
Moreover, this analysis favours an increase of the $R_s$ ratio.
\item If we neglect exchange and penguin annihilation topologies -- the most
recent bounds from the  $B_d\to K^+K^-$ and $B_s\to\pi^+\pi^-$ data do 
not indicate any anomalous enhancement -- we obtain an interesting interplay
between the $B_d\to\pi^+\pi^-$, $B_s\to K^+K^-$ and $B_d\to\pi^\mp K^\pm$,
$B_s\to\pi^\pm K^\mp$ systems. This allows us to resolve the ambiguity
in the extraction of $\gamma$ from the former decays, as noted above, and
to determine an $SU(3)$-breaking form-factor ratio from the data, which agrees 
with the result of a recent QCD sum-rule calculation used in our analysis, 
and disfavours large non-factorizable effects. Moreover, we can also make 
predictions for $\mbox{BR}(B_s\to \pi^\pm K^\mp)$, which point towards an 
increase with respect to the current CDF central value.
\end{itemize}
The $U$-spin extraction of $\gamma$ from the $B_d\to\pi^+\pi^-$, $B_s\to K^+K^-$ 
system is already for the first Tevatron data one of the most accurate determinations
on the market, and can be subsequently further optimized. In our analysis, we 
obtain a remarkable agreement with the SM picture of CP violation. Thanks to 
the start of the LHC, we will soon enter a new era for the exploration of the 
$B_s$-meson system. The LHCb experiment will then allow us to obtain a much
sharper picture of the strategy discussed in this paper and to exploit its full
physics potential. Moreover, also precision measurements of $\gamma$ from 
tree-level processes will become possible, which are another -- still missing -- 
key element for the search of NP. It will be very interesting to compare all 
these measurements with one another and to confront the 
Kobayashi--Maskawa mechanism of CP violation with another round of stringent tests.

\end{document}